%% file: AMS_SA_draft.tex
\documentclass[11pt,a4paper]{article}

\pdfoutput=1

\usepackage{graphicx}
\usepackage{amsmath,amssymb,cite}
\usepackage{authblk}
\usepackage{geometry}
\usepackage{tikz}
\usetikzlibrary{positioning,arrows,patterns}
\usetikzlibrary{decorations.pathmorphing}
\usetikzlibrary{decorations.markings}
\usetikzlibrary{calc}
\tikzset{
    photon/.style={decorate, decoration={snake}, draw=black, thick},
    fermionnoarrow/.style={draw=black, postaction={decorate}, thick},
    scalar/.style={draw=black, postaction={decorate}, thick, dashed},
    fermion/.style={draw=black, postaction={decorate},decoration={markings,mark=at position .55 with {\arrow{>}}}, thick},
    gluon/.style={decorate, draw=black, decoration={coil,amplitude=4pt, segment length=5pt}, thick},
    vertex/.style={draw,shape=circle,fill=black,minimum size=3pt,inner sep=0pt} 
}

\graphicspath{{./}{./Figures/}}

\newcommand{\order}[1]{\mathcal{O}(#1)}
\newcommand{\expect}[1]{\left\langle #1 \right\rangle}
\newcommand{\set}[1]{\mathbb{#1}}
\newcommand{\abs}[1]{\left\lvert #1 \right\rvert}

\newcommand{\modeqref}[1]{eq.~\eqref{#1}}
\newcommand{\Modeqref}[1]{Eq.~\eqref{#1}}
\newcommand{\modeqsref}[1]{eqs.~\eqref{#1}}
\newcommand{\figref}[1]{fig.~\ref{#1}}

\newcommand{\tabref}[1]{table~\ref{#1}}

\newcommand{\secref}[1]{section~\ref{#1}}
\newcommand{\secsref}[1]{sections~\ref{#1}}
\newcommand{\appref}[1]{appendix~\ref{#1}}
\newcommand{\refcite}[1]{ref.~\cite{#1}}
\newcommand{\Refcite}[1]{Ref.~\cite{#1}}
\newcommand{\refscite}[1]{refs.~\cite{#1}}

\newcommand{\dsym}{\mathcal{D}}
\newcommand{\lag}{\mathcal{L}}

\begin{document}

\title{Low-Temperature Enhancement of Semi-annihilation \\and the AMS-02 Positron Anomaly}
\author[1]{Yi Cai\thanks{\texttt{caiyi.pku@gmail.com}}} 
\author[2]{Andrew Spray\thanks{\texttt{a.spray.work@gmail.com}}}
\affil[1]{School of Physics, Sun Yat-sen University, Guangzhou, 510275, China}
\affil[2]{Center for Theoretical Physics of the Universe, Institute for Basic Science (IBS), Daejeon, 34126, Korea}

\maketitle

\begin{abstract}
	Semi-annihilation is a generic feature of particle dark matter that is most easily probed by cosmic ray experiments.  We explore models where the semi-annihilation cross section is enhanced at late times and low temperatures by the presence of an $s$-channel resonance near threshold.  The relic density is then sensitive to the evolution of the dark matter temperature, and we compute expressions for the associated Boltzmann equation valid in general semi-annihilating models.  At late times, a \emph{self-heating} effect warms the dark matter, allowing number-changing processes to remain effective long after kinetic decoupling of the dark and visible sectors.  This allows the semi-annihilation signal today to be enhanced by up to five orders of magnitude over the thermal relic cross section.  As a case study, we apply this to a dark matter explanation of the positron excess seen by AMS-02.  We see that unlike annihilating dark matter, our model has no difficulty fitting the data while also giving the correct relic density.  However, constraints from the CMB and $\gamma$-rays from the galactic centre do restrict the preferred regions of parameter space.
\end{abstract}

\input{./Files/Intro}


\input{./Files/SAReview}


\input{./Files/TempEvo}


\input{./Files/BW}


\input{./Files/model}


\input{./Files/amsfit}


\input{./Files/rdens}


\input{./Files/conc}

\section*{Acknowledgements}

We thank Yu Gao for the discussion of positron propagation in the galaxy.  
This work was supported by IBS under the project code, IBS-R018-D1.  AS thanks Sun Yat-sen University for its hospitality while this work was being completed.

\appendix

\input{./Files/Derivation}

\input{./Files/UV}

\bibliography{SArefs}{}
\bibliographystyle{JHEP}

\end{document}

%% file: Files/Intro.tex
\section{Introduction}\label{sec:intro}

The dark matter problem remains perhaps the most compelling single piece of evidence for the existence of new particles.  Thermal freeze-out further provides a simple mechanism to explain the observed relic density with only one or a few states, and can be easily motivated on particle physics grounds.  However, substantial experimental progress in recent years has covered much of the parameter space of traditional frameworks such as supersymmetry.  In such an environment, it is important to check that the assumptions in conventional approaches are not too severe; and any generic weakening of the present bounds would motivate closer study.

\begin{figure}
	\centering
	\begin{tikzpicture}[node distance=1cm and 1.75cm]
		\coordinate (v1);
		\coordinate[above left = of v1, label=above left:$\chi$] (i1);
		\coordinate[below left = of v1, label=below left:$\chi$] (i2);
		\coordinate[above right = of v1, label=above right:{$V$}] (o1);
		\coordinate[below right = of v1, label=below right:{$V$}] (o2);
		\draw[fermionnoarrow] (i1) -- (v1);
		\draw[fermionnoarrow] (v1) -- (i2);
		\draw[fermionnoarrow] (o2) -- (v1);
		\draw[fermionnoarrow] (v1) -- (o1);
		\draw[fill = white] (v1) circle (1);
		\fill[pattern = north west lines] (v1) circle (1);
	\end{tikzpicture}\qquad\qquad
	\begin{tikzpicture}[node distance=1cm and 1.75cm]
		\coordinate (v1);
		\coordinate[above left = of v1, label=above left:$\chi$] (i1);
		\coordinate[below left = of v1, label=below left:$\chi$] (i2);
		\coordinate[above right = of v1, label=above right:{$\chi$}] (o1);
		\coordinate[below right = of v1, label=below right:{$V$}] (o2);
		\draw[fermionnoarrow] (i1) -- (v1);
		\draw[fermionnoarrow] (v1) -- (i2);
		\draw[fermionnoarrow] (o2) -- (v1);
		\draw[fermionnoarrow] (v1) -- (o1);
		\draw[fill = white] (v1) circle (1);
		\fill[pattern = north west lines] (v1) circle (1);
	\end{tikzpicture}
	\caption{Two types of dark sector-visible sector interactions, where $\chi$ ($V$) is any dark (visible) field.  (Left): DM annihilation to/from, or scattering off, the SM; this is the only possibility when the DM is stabilised by a $\set{Z}_2$ symmetry.  (Right): Semi-annihilation, a non-decay process with an odd number of external dark fields, generically possible when the stabilising symmetry is not a $\set{Z}_2$.}\label{fig:SA}
\end{figure}
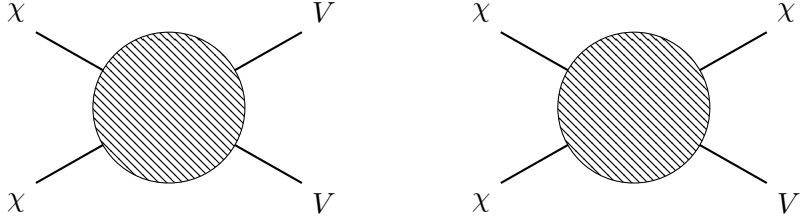

One such exception is given by semi-annihilation~\cite{1003.5912} (SA), non-decay processes with an odd number of external dark states.  Models of particle dark matter (DM) almost always require a symmetry, either exact or (for decaying DM) very weakly broken, under which the dark sector is charged and the Standard Model (SM) is neutral.  In the most commonly-studied scenarios, that global symmetry is a single $\set{Z}_2$.  The immediate implication is that all processes must have an even number of external dark sector particles, and the dominant number-changing processes are $2\to 2$ (co)-annihilation, see the left side of \figref{fig:SA}.  Crossing symmetry then relates the relic density to signal rates at colliders or direct detection experiments, and it is the combination of all these searches that makes bounds so severe.  However, \emph{any other} global symmetry may allow SA, illustrated in the right side of \figref{fig:SA}, which can be important for the determination of the thermal relic density while irrelevant at colliders and direct detection experiments.  Semi-annihilating dark matter (SADM) is then both generic \emph{and} less constrained than conventional models.

Though constraints are weakened in SADM, they are not absent.  The diagram in the left of \figref{fig:SA} can not be forbidden by any symmetry, so will always be present; and SA itself can lead to indirect signals in cosmic ray experiments~\cite{Zeldovich:1980st}.  It is therefore important to explore the model space and discover to what extent current constraints do apply.  \Refcite{1611.09360} made an initial step in this direction by systematically constructing effective field theories describing all possible $2\to 2$ SA with SM final states.  However, that approach implicitly assumed that the low-energy interactions between the dark and visible sectors are non-perturbative with simple velocity dependence.  Exceptions to this behaviour are well-known, and include the Sommerfeld effect~\cite{hep-ph/0307216,1005.4678}, bound state formation~\cite{0812.0559,1407.7874,1706.01894,1801.05821}, and the presence of an $s$-channel resonance near threshold~\cite{0812.0072}.  All of these have been studied in the context of DM both for how they affect the relic density, and also for the possibility that they can lead to enhanced cross sections today.  As indirect detection is the most robust search channel for SADM, it is worth considering the interplay of these two aspects of phenomenology.  For previous work in this direction see~\cite{1509.08481,1706.09974}.

In \refcite{1705.10777} it was emphasised that when annihilation is enhanced at low temperature, the final relic density depends sensitively on the DM temperature $T_\chi$.  It is usually assumed that the DM remains in kinetic equilibrium, $T_\chi = T_{SM} \equiv T$, till all number changing process cease to be important.  While a safe assumption in conventional scenarios which freeze-out at relatively high temperatures, $T_\chi/m_\chi \sim 25$, enhanced cross sections can continue to be relevant till much later, $T_\chi/m_\chi \gtrsim 10^{3}$.   This is especially important in the case of an $s$-channel resonance; the DM-SM scattering process related by crossing symmetry will be much smaller, leading to earlier kinetic decoupling of the dark and visible sectors.

The temperature evolution of DM has attracted recent attention in a number of contexts~\cite{1706.07433,Bringmann:2006mu,1603.04884,1705.10777,1803.08062,1803.07518,1711.02970,1710.06447,1709.09717}.  The temperature of SADM specifically was previously studied in \refscite{1707.09238,1805.05648}; however, those works focused on a specific type of single-component SA, and provided expressions valid only for Maxwellian phase space distributions.  We extend their work to SADM phenomenology of the type discussed in \refcite{1611.09360}, provide expressions valid for arbitrary phase space distributions, and derive new approximations useful in the Maxwellian non-relativistic case.  Doing so illustrates the generality of the earlier result that post-kinetic decoupling, SADM redshifts as \emph{radiation}, $T_\chi \propto T$, and not matter, $T_\chi \propto T^2$.  This is despite the DM remaining strongly non-relativistic, and is instead due to a \emph{self-heating} effect where SA converts mass energy to dark sector kinetic energy.  For sufficiently large SA cross sections, the DM can even be \emph{hotter} than the visible sector prior to matter-radiation equality.  The implications are obvious: if kinetic decoupling occurs before freeze-out of low-temperature enhanced cross sections, since SADM is \emph{warmer} than conventional DM it will have \emph{smaller} number changing rates.  The observed relic density can be obtained for larger DM-SM couplings, leading to larger potential signals today.  

To illustrate this effect, we consider a concrete example: a SADM model for the positron excess seen by PAMELA~\cite{0810.4995} and AMS-02~\cite{Aguilar:2013qda,Aguilar:2014mma,Zimmermann:2017abs}.  These experiments measure the cosmic ray positron flux for 10\,GeV\,$\lesssim E \lesssim$\,1\,TeV to be orders of magnitude larger than the expected secondary backgrounds~\cite{astro-ph/9710124,astro-ph/9808243}.  While astrophysical explanations exist, most notably millisecond Pulsars~\cite{0810.1527,0903.1310,1108.4827}, this observation has also proved fertile ground for DM model building, \emph{e.g.}~\cite{0903.0122,1707.09313,1306.2959,1409.4590,1403.1212,1508.02881} and references therein (see \refcite{1706.09974} for an earlier model involving SADM)\footnote{See also \refcite{1706.02336} for a model of SADM that attempts to fit the AMS-02 anti-proton and anti-Helium data.}.  These models require (semi)-annihilation cross sections approximately three orders of magnitude larger than the thermal relic cross section.  However, it was shown in \refcite{1106.6027} that the \emph{theoretical} maximum enhancement possible with a Breit-Wigner resonance is only $\lesssim 10^2$.  This limit arises precisely because the temperature drop after kinetic decoupling leads to a suppression in the relic density, bounding the annihilation cross sections from above.  The higher post-decoupling temperature of SADM is then a natural avenue to circumvent this obstacle.

The outline of this paper is as follows.  We first review some details on semi-annihilation, and in particular the concept of dark partners, in \secref{sec:DP}.  In \secref{sec:TempEvo} we review how the dark matter temperature $T_\chi$ is defined and evolves, expand on previous results for SADM, and derive its asymptotic behaviour.  In \secref{sec:bw}, we discuss cross sections enhanced by an $s$-channel resonance, how they depend on the dark matter temperature evolution, and the role that SA can play.  We then perform the positron excess case study, first constructing an explicit model in \secref{sec:model}, then fitting the putative signal in \secref{sec:fit}, and finally demonstrating that our model can simultaneously fit the signal cross section and thermal relic density in \secref{sec:rd}.  We end with our conclusions in \secref{sec:conc}.  Some additional aspects are deferred to the appendices; \appref{app:fullBE} contains more technical details on the Boltzmann equations and thermally averaged cross sections, and \appref{app:UV} outlines a lepton-number motivation for our simplified model.

%% file: Files/SAReview.tex
\section{Semi-annihilation and dark partners}\label{sec:DP}

We assume the existence of an unbroken global symmetry group $\dsym \neq \set{Z}_2$, under which all SM fields are neutral.  We refer to the states (un)charged under $\dsym$ as the dark (visible) sector, which include at least the dark matter $\chi$ (SM).  Conservation of $\dsym$ means that SA has at least three external dark sector particles, with the minimal possibility being $\chi\chi \to \chi^\dagger + SM$.  The DM must be neutral under the unbroken SM gauge group; the same must then hold for the set of SM particles in the final state.  For $2\to 2$ processes, the only possible SM final states are then the photon, $Z$, Higgs and neutrinos.  However, as discussed in \refcite{1611.09360}, only the Higgs is possible in renormalisable minimal models, with the other states demanding multi-component dark sectors.

If we want to allow other SM final states, we must expand our interest to at least $2\to 3$ processes.  In general, these processes will be phase-space suppressed, making it harder for them to be relevant for thermal freeze out.  However, we can circumvent that suppression by introducing additional unstable states, such that the SA is a $2\to 2$ process followed by one or more decays.  One commonly studied option uses additional \emph{visible sector} states $\phi$; the relevant channel is then $\chi\chi \to \chi^\dagger\phi$, followed by $\phi$ decay to the SM.  For examples, see \refscite{1003.5912,1805.05648,1707.09238,1706.09974,1407.5492}.

In this work, we will follow an alternative option introduced in \refcite{1611.09360}, where we add further \emph{dark sector} particles $\Psi$, charged under both $\dsym$ and the SM gauge group.  In these models, SA is $\chi\chi \to \Psi^\dagger V$, where $V$ is an SM state, followed by $\Psi$ decay.  In order for SA to be relevant during freeze-out, $m_\Psi < 2m_\chi$.  The only possible decay channel for $\Psi$ is $\Psi \to \chi + SM$, which requires $m_\Psi > m_\chi$ and that $\dsym$ is (equivalent to) $\set{Z}_3$.  Finally, in order for SA to be a $2\to 2$ process, $\Psi$ must have the same SM quantum numbers as $V$, for which reason we call these states \emph{dark partners}.

One important complication in models which contain dark partners is that the Boltzmann equation for the dark matter contains an explicit dependence on the number density of $\Psi$.  Both the SA term itself as well as the $\Psi$ decay term depend on $n_\Psi$ and are relevant to dark matter freeze-out.  This means that in general it is necessary to integrate the coupled equations for both $Y_{\chi,\Psi} = n_{\chi,\Psi}/s$, with $s$ the entropy density, when determining the relic abundance of $\chi$.  We discuss these expressions more fully in \secref{subsec:TESA} and \appref{app:fullBE}.  Here we outline the general evolution of $Y_\Psi$, its asymptotic behaviour and approximate solution after $\Psi$ annihilations freeze out, and a useful simplification when $\Psi$ decays promptly.

The Boltzmann equation for the dark partner is\footnote{This is true if $\chi$-SM and $\Psi$-SM scattering remain efficient.  We relax the former assumption later.}
\begin{align}
	\frac{dY_\Psi}{dx} & = - \frac{\Gamma_\Psi}{xHZ} \, \biggl( Y_\Psi - Y_\chi \, \frac{Y_\Psi^{eq}}{Y_\chi^{eq}} \biggr) + \frac{1}{2} \, \frac{s}{x H Z} \, \langle \sigma v (\chi \chi \to \Psi^\dagger V) \rangle \, \biggl( Y_\chi^2 - Y_\Psi \, \frac{(Y_\chi^{eq})^2}{Y_\Psi^{eq}} \biggr) \notag \\
	& \quad - \frac{s}{x H Z} \, \langle \sigma v (\Psi^\dagger \Psi \to SM) \rangle \, \bigl( Y_\Psi^2 - (Y_\Psi^{eq})^2 \bigr) + \ldots\,. \label{eq:dYPsi}
\end{align}
The factor of one-half is a symmetry factor associated with identical particles.  We have introduced the usual inverse temperature $x = m_\chi/T$ and 
\begin{equation}
	Z = \biggl( 1 - \frac{x}{3 g_{\ast S}} \, \frac{d g_{\ast S}}{dx} \biggr)^{-1} \,,
\end{equation}
where $g_{\ast S}$ is the effective number of relativistic degrees of freedom, such that the entropy density $s = 2\pi^2 g_{\ast S} T^3/45$.  The dots in \modeqref{eq:dYPsi} denote contributions from co-annihilation $\Psi \chi^\dagger \to SM$ and semi-co-annihilation $\Psi\chi \to \chi^\dagger V$; these are qualitatively similar to annihilation, in that the reverse channels are exponentially Maxwell-suppressed at low temperatures.  These three processes all enforce $Y_\Psi = Y_\Psi^{eq}$ at early times but freeze out in the usual manner. 

In contrast, (inverse) decay processes become \emph{more} relevant at late times.  Prior to $\chi$ freeze out\footnote{We assume that since $m_\Psi > m_\chi$, dark partner annihilations freeze out first.  If $\Psi$ has SM gauge interactions, its annihilation cross section can be larger than for DM and this might not hold.}, $Y_\chi \approx Y_\chi^{eq}$ and the decay term drives $Y_\Psi$ towards equilibrium.  It will dominate expansion for $x > x^{dec}$ where
\begin{equation}
	x^{dec} \, H(x^{dec}) = \Gamma_\Psi \,, \qquad\qquad x^{dec} \sim \frac{10^{-15}}{\Gamma_\Psi/m_\Psi} \, \biggl( \frac{m_\chi}{1 \, \text{TeV}} \biggr) \,.
\label{eq:defxdec}\end{equation}
It follows that inverse decays will maintain $\Psi$ equilibrium for all but the longest-lived dark partners till $Y_\chi$ appreciably differs from its equilibrium value.  After $\chi$ freeze out, we can approximate \modeqref{eq:dYPsi} by neglecting all terms proportional to at least one power of $Y_i^{eq}$.  We can also usually neglect terms quadratic in $Y_\Psi \ll 1$; in particular, the decay term will also dominate over the annihilation term provided that
\begin{equation}
	\frac{\Gamma_\Psi}{m_\Psi} > \langle \sigma v (\bar{\Psi}\Psi \to SM) \rangle \, \frac{n_\Psi}{m_\Psi} \sim 10^{-17} \, \biggl( \frac{\langle \sigma v (\bar{\Psi}\Psi \to SM) \rangle}{3 \times 10^{-26} \, \text{cm}^3 \, \text{s}^{-1}} \biggr) \, \biggl( \frac{n_\Psi}{n_\Psi^{eq} (T = m_\Psi/25)} \biggr) \,,
\end{equation}
which is typically a weaker condition than \modeqref{eq:defxdec}.  We then have the approximate Boltzmann equation
\begin{equation}
	\frac{dY_\Psi}{dx} \approx - \frac{\Gamma_\Psi}{xHZ} \, \biggl( Y_\Psi - Y_\chi \, \frac{Y_\Psi^{eq}}{Y_\chi^{eq}} \biggr) + \frac{1}{2} \, \frac{s}{x H Z} \, \langle \sigma v (\chi \chi \to \Psi^\dagger V) \rangle \, Y_\chi^2 \,.
\label{eq:dYPsiapprox}\end{equation}
The inverse decay term is also exponentially small at low temperatures.  However, it is the least such suppressed term, and can be relevant at intermediate times.  The approximate solution for $Y_\Psi$ is 
\begin{equation}
	Y_\Psi \approx Y_\chi \, \frac{Y_\Psi^{eq}}{Y_\chi^{eq}} + \frac{s \, \langle \sigma v (\chi \chi \to \Psi^\dagger V) \rangle}{2 \Gamma_\Psi} \, Y_\chi^2 \,.
\label{eq:YPsiAppr2}\end{equation}
This can also describe the number density at early times when the width is sufficiently large, since prior to $\chi$ freeze-out the first term gives $Y_\Psi = Y_\Psi^{eq}$.  The condition for this to be true is given by
\begin{equation}
	\frac{\Gamma_\Psi}{m_\Psi} > 10^{-3} \, x^{-3/2} \, e^{-(2-m_\Psi/m_\chi)x} \, \biggl( \frac{m_\chi}{m_\Psi} \biggr)^{5/2} \biggl( \frac{\langle \sigma v (\chi \chi \to \Psi^\dagger V) \rangle}{3 \times 10^{-26} \, \text{cm}^3 \, \text{s}^{-1}} \biggr) \biggl( \frac{m_\chi}{1\,\text{TeV}} \biggr)^2 \,.\label{eq:LargePsiDecay}
\end{equation}
Thanks to the exponential suppression, this will generally hold for any unsuppressed two-body decay for $x \gtrsim 1$.

The history of the usual dark partner evolution is then one where it starts in equilibrium till $\chi$ freeze out, after which \modeqref{eq:YPsiAppr2} is true.  When the first term dominates, the dark partner number density is set by inverse decay; it is larger than the equilibrium density and proportional to $Y_\chi$.  However, that term is exponentially suppressed and so at sufficiently low temperatures the SA term will always come to dominate.  The asymptotic solution for $Y_\Psi$ is 
\begin{equation}
	\lim_{x\to\infty} Y_\Psi \approx \frac{s \, \langle \sigma v (\chi \chi \to \Psi^\dagger V) \rangle_{T=0}}{2 \Gamma_\Psi} \, Y_\chi^2 \,.
\label{eq:YPsiInf}\end{equation}
Physically, since SA production of $\Psi$ is unique in not being Maxwell-suppressed and $Y_\chi$ asymptotes to a constant, it is the only relevant production method of dark partners at late times.  The number density of $\Psi$ is then set by balancing this process with the decay rate.

If the approximations leading to \modeqref{eq:YPsiAppr2} hold, we can use this to simplify our computation of the relic density.  In particular, if the decay term is prompt and dominant during $\chi$ freeze out, then we can rewrite the contribution of SA and $\Psi$ decay to the $Y_\chi$ Boltzmann equation as
\begin{align}
	\frac{dY_\chi}{dx} & = \frac{\Gamma_\Psi}{xHZ} \, \biggl( Y_\Psi - Y_\chi \, \frac{Y_\Psi^{eq}}{Y_\chi^{eq}} \biggr) - \frac{s}{x H Z} \, \langle \sigma v (\chi \chi \to \Psi^\dagger V) \rangle \, \biggl( Y_\chi^2 - Y_\Psi \, \frac{(Y_\chi^{eq})^2}{Y_\Psi^{eq}} \biggr) + \ldots \notag \\
	& \approx - \frac{1}{2} \, \frac{s}{x H Z} \, \langle \sigma v (\chi \chi \to \Psi^\dagger V) \rangle \, Y_\chi \bigl( Y_\chi - Y_\chi^{eq} \bigr) + \ldots
\label{eq:dYchiAppr}\end{align}
In the second line, we used \modeqref{eq:dYPsi} to eliminate the decay term, substituted the first term in \modeqref{eq:YPsiAppr2} for $Y_\Psi$, and neglected $dY_\Psi/dx$ since $Y_\Psi \ll Y_\chi$.  More generally, we could use both terms in \modeqref{eq:YPsiAppr2}, but the relevance of \modeqref{eq:dYchiAppr} is that it has the usual semi-annihilation form for theories \emph{without} dark partners.  Physically, if the dark partner decays promptly during $\chi$ freeze-out, we can approximate the combined SA plus decay process as a single SA event, $\chi\chi \to \chi^\dagger + SM$.  The factor of one-half means that the cross section required for a thermal relic through SA alone is enhanced over the usual canonical value.  There is an additional factor of 2 due to SADM always being complex, so the canonical thermal SA cross section is $\langle \sigma v \rangle_0 = 1.2 \times 10^{-25}\,$cm$^3\,$s$^{-1}$.

%% file: Files/TempEvo.tex
\section{Temperature evolution  of semi-annihilating dark matter}\label{sec:TempEvo}

The details of the temperature evolution of dark matter, and of kinetic decoupling from the SM, have recently been discussed in a number of contexts~\cite{1706.07433,Bringmann:2006mu,1707.09238,1603.04884,1705.10777,1805.05648,1803.08062,1803.07518,1711.02970,1710.06447,1709.09717,phdscatter}.  How this behaviour changes in the presence of semi-annihilation was first explored in \refscite{1707.09238,1805.05648}; we expand on and provide alternative forms for their results.  In this section, we first review the definition of the dark matter temperature, the consequences of it differing from the SM plasma temperature, and its Boltzmann equation.  We then give expressions that correspond to the contributions of SA and dark partner decay.  In both cases we assume $CP$ is conserved.  Finally, we discuss the distinct asymptotic behaviour of the temperature in the presence of SA, first described in \refcite{1707.09238}, where the temperature redshifts like radiation due to a self-heating effect.  We clarify how this behaviour is generic, and under what circumstances it can lead to dark matter which is asymptotically hotter or colder than the SM.

\subsection{Review}\label{subsec:TEreview}

We follow \refcite{1706.07433} in defining the dark matter temperature $T_\chi$ and a dimensionless variable $y_\chi$ as
\begin{equation}
	T_\chi = \frac{g_\chi}{3n_\chi} \int \frac{d^3p_\chi}{(2\pi)^3} \, \frac{\vec{p}_\chi^{\,2}}{E_\chi} \, f_\chi (p_\chi) \,, \qquad y_\chi = \frac{m_\chi T_\chi}{s^{2/3}} \,,
\label{eq:defTchi}\end{equation}
where $f_\chi (p_\chi)$ is the dark matter phase space distribution, $g_\chi$ the number of internal degrees of freedom, and $s$ the entropy density.  The former expression is defined for any distribution function, but produces the expected result if the phase space function is Maxwellian, $f_\chi \propto e^{-E_\chi/T_\chi}$.  The dimensionless parameter is so defined as, for normal dark matter, it is approximately constant after kinetic decoupling.  As discussed below this is no longer true for SADM, but we retain the same definition to maintain consistency with and make use of the literature.

The first impact of allowing $T_\chi \neq T_{SM}$ is that the thermal averages in the number density Boltzmann equation are modified.  Specifically, forward annihilation must be evaluated using the actual functions $f_\chi$, while the inverse rates use the equilibrium distributions $f_\chi^{eq}$:
\begin{equation}
	\frac{dY_\chi}{dx} \supset - \frac{s}{xHZ} \, \Bigl( Y_\chi^2  \, \langle \sigma v (\chi\chi^\dagger \to SM) \rangle_{neq} - (Y_\chi^{eq})^2 \, \langle \sigma v (\chi\chi^\dagger \to SM) \rangle_{eq} \Bigr) \,.
\label{eq:dYwithTchi}\end{equation}
The thermally averaged cross sections are defined in the usual way, see \modeqref{eq:svneq} in \appref{app:fullBE} for more details.  More importantly, we must include an additional Boltzmann equation for $y_\chi$ that describes the evolution of the DM temperature.  The general form is~\cite{1706.07433}
\begin{equation}
	\frac{1}{y_\chi} \, \frac{dy_\chi}{dx} = - \frac{1}{Y_\chi} \, \frac{dY_\chi}{dx} + \frac{1}{x Z} \, \frac{1}{3T} \, \biggl\langle \frac{p^4}{E^3} \biggr\rangle_{T_\chi} + \frac{m_\chi}{xHZ} \, \sum_i C_2^{(i)}
\label{eq:dytgen}\end{equation}
where the $C_2^{(i)}$ are moments of collision operators, and the sum includes all terms that contribute to the $Y_\chi$ Boltzmann equation plus scattering processes between DM and the SM.  For annihilation-like terms, they are given by a simple extension of \modeqref{eq:dYwithTchi},
\begin{equation}
	C_2^{(ann)} = - \frac{s}{m_\chi} \, \biggl( Y_\chi \, \langle \sigma v (\chi\chi^\dagger \to SM) \rangle_{2,neq} - \frac{(Y_\chi^{eq})^2}{Y_\chi} \, \langle \sigma v (\chi\chi^\dagger \to SM) \rangle_{2,eq} \biggr) 
\label{eq:CollTAnn}\end{equation}
where the modified thermal averages are
\begin{equation}
	 \langle \sigma v (\chi\chi^\dagger \to SM) \rangle_{2,neq} = \frac{g_\chi^2}{3 n_\chi^2 T_\chi} \int \frac{d^3 p_\chi}{(2\pi)^3} \, \frac{d^3 p_{\chi^\dagger}}{(2\pi)^3} \, \frac{\vec{p}_\chi^{\,2}}{E_\chi} \, f_\chi (p_\chi) \, f_\chi (p_{\chi^\dagger}) \, \sigma \, \frac{\sqrt{(p_\chi \cdot p_{\chi^\dagger})^2 - m_\chi^4}}{E_\chi \, E_{\chi^\dagger}} \,.
\label{eq:AnnytTA}\end{equation}
The equilibrium terms have the obvious replacement $f_\chi \to f_\chi^{eq}$.  For pure $s$-wave cross sections, $\langle\sigma v\rangle_2 = \langle\sigma v\rangle$, but the two functions differ when higher partial waves are relevant.  For the scattering terms, we have instead
\begin{equation}
	C_2^{(sca)} = \frac{\gamma(T)}{m_\chi} \, \biggl( \frac{y_{eq}}{y_\chi} - 1 \biggr)
\label{eq:C2scatter}\end{equation}
where $y_{eq}$ is $y_\chi$ when $T_\chi = T$ and the $\gamma$-function is~\cite{phdscatter}
\begin{equation}
	\gamma (T) = \frac{g_{SM}}{3\pi^2 m_\chi^2 T} \int d\omega \, f_{SM} (\omega) \bigl( 1 \pm f_{SM} (\omega) \bigr) k^4 \int d\Omega \, (1 - \cos\theta) \, \frac{d\sigma}{d\Omega} \,,
\label{eq:gammascatter}\end{equation}
with the scattering cross section $\sigma$ evaluated at centre of mass energy $E_{cm}^2 = m_\chi^2 + 2 m_\chi \, \omega$.

\subsection{Semi-annihilation and kinetic decoupling}\label{subsec:TESA}

In light of our discussion in \secref{sec:DP}, SADM models can be divided into two classes: those with and without dark partners.  The latter case was previously discussed in \refscite{1707.09238,1805.05648}, but we expand on their discussion.  The contribution of the forward process to both the number density and temperature Boltzmann equations have the same form as in \modeqsref{eq:dYwithTchi} and~\eqref{eq:CollTAnn}, replacing the annihilation cross section with the SA one.  The inverse process is more complicated; the temperature difference prevents us from being able to factor the collision operator integrals in terms of the forward cross section as usual.  The best exact expression uses the inverse cross section directly, so that 
\begin{equation}
	\frac{dY_\chi}{dx} \supset - \frac{1}{2} \, \frac{s}{xHZ} \, Y_\chi \, \Bigl( Y_\chi \, \langle \sigma v (\chi\chi \to \chi^\dagger + SM) \rangle_{neq} - Y_{SM}^{eq} \, \langle \sigma v (\chi^\dagger + SM \to \chi\chi) \rangle_{T_\chi,T} \Bigr) \,,
\label{eq:dYSA1Tchi}\end{equation}
where the factor of one-half is the usual result of SA changing $\chi$ number by one instead of two, and the thermal average of the inverse process is 
\begin{align}
	\langle \sigma v (\chi^\dagger + SM \to \chi\chi) \rangle_{T_\chi,T} & = \frac{g_\chi \, g_{SM}}{n_\chi \, n_{SM}^{eq}} \int \frac{d^3 p_{\chi^\dagger}}{(2\pi)^3} \, \frac{d^3 p_{SM}}{(2\pi)^3} \, f_\chi (p_{\chi^\dagger}) \, f_{SM} (p_{SM}) \notag \\*
	& \quad \times \sigma (\chi^\dagger + SM \to \chi\chi) \, \frac{\sqrt{(p_{\chi^\dagger} \cdot p_{SM})^2 - m_\chi^2 m_{SM}^2}}{E_{\chi^\dagger} \, E_{SM}} \,. \label{eq:InvSVavg}
\end{align}
We have written this assuming a single SM particle, but the generalisation to more is conceptually simple.  Even in a $2\to 2$ process, however, \modeqref{eq:InvSVavg} is awkward to compute, being a function of two inputs ($T$ and $T_\chi$) that requires at least a two-dimensional numerical integration.  To this end, we note a useful simplification that occurs in the non-relativistic limit.  First we rewrite the phase space functions as
\begin{equation}
	f_\chi (p_{\chi^\dagger}) \, f_{SM} (p_{SM}) = \frac{n_\chi}{n_\chi^{eq} (T_\chi)} \, e^{-E_{\chi^\dagger}/T_\chi} \, e^{-E_{SM}/T} = \frac{n_\chi}{n_\chi^{eq} (T_\chi)} \, e^{-(E_{\chi_1} + E_{\chi_2})/T} \, e^{-E_{\chi^\dagger} \, \Delta_\chi/m_\chi} \,,
\label{eq:AltPhSp}\end{equation}
where we have defined the difference of inverse temperatures,
\begin{equation}
	\Delta_\chi = \frac{m_\chi}{T_\chi} - \frac{m_\chi}{T} \,.
\label{eq:defDelta}\end{equation}
Next, we note that the typical momentum of the SA final state $\chi^\dagger$ is large compared to the $\chi$ thermal momentum $p_{\chi,th} \sim T$. For a $2\to 2$ process and neglecting the SM mass, $\lvert \vec{p}_{\chi^\dagger}\rvert = 3m_\chi/4$; including visible sector masses or extra final state particles will result in lower momenta but still typically $\mathcal{O}(m_\chi)$.  It follows that the boost from the cosmological frame to the collision centre of mass frame will be a correction of $\mathcal{O} (x^{-1})$.  In many cases, including those studied in this paper, scattering is more important till $x \gtrsim 100$ so this is a negligible sub-percent effect.  Then we may define a more easily-computable function
\begin{equation}
	\mathcal{S} (T, T_\chi) = \frac{n_\chi^{eq} (T)}{n_\chi^{eq} (T_\chi)} \, \frac{1}{\sigma (\chi\chi \to \chi^\dagger SM)} \, \frac{1}{2E_{cm}^2 (1 - 4m_\chi^2/E_{cm}^2)^{1/2}} \int d\Pi_n \, \sum \lvert \mathcal{M} \rvert^2 \, e^{-E_{\chi^\dagger} \Delta_\chi/m_\chi} \,,
\label{eq:defSTT}\end{equation}
with $d\Pi_n$ the $n$-body Lorentz-invariant phase space (including momentum-conserving delta funciton) and $E_{cm}$ the total centre-of-mass frame energy, such that $\mathcal{S} (T, T) = 1$ and
\begin{equation}
	\frac{dY_\chi}{dx} \supset - \frac{1}{2} \, \frac{s}{xHZ} \, Y_\chi \, \Bigl( Y_\chi \, \langle \sigma v (\chi\chi \to \chi^\dagger + SM) \rangle_{neq} - Y_{\chi}^{eq} \, \langle \sigma v (\chi\chi \to \chi^\dagger + SM) \rangle_{eq} \, \mathcal{S} (T, T_\chi) \Bigr) \,.
\end{equation}
Physically, $\mathcal{S}$ is the mean value of the exponential over all energies the $\chi^\dagger$ has (multiplied by the ratio of equilibrium densities).  It follows that the integral in \modeqref{eq:defSTT} is trivial for $2 \to 2$ processes, and more generally can be reduced to a one-dimensional integral as a function of one parameter, $\Delta_\chi$.  In the non-relativistic limit, $n_\chi^{eq} (T)/n_\chi^{eq} (T_\chi) = (T/T_\chi)^{3/2} \, e^{\Delta_\chi}$, which reduces the exponential dependence of $\mathcal{S}$ on $\Delta_\chi$.  For the usual case $T_\chi \leq T$, $\mathcal{S} \leq 1$ is a suppression factor.

The contribution of this type of SA to the $y_\chi$ Boltzmann equation follows a similar pattern.  We focus on the non-relativistic approximation for Maxwellian phase space distributions.  We define an analogous function to \modeqref{eq:defSTT}, 
\begin{equation}
	\mathcal{S}_{\mathcal{T}} (T, T_\chi) = \frac{n_\chi^{eq} (T)}{n_\chi^{eq} (T_\chi)} \, \frac{1}{\sigma} \, \frac{1}{\mathcal{U}^S_\chi} \, \frac{1}{2E_{cm}^2 (1 - 4m_\chi^2/E_{cm}^2)^{1/2}} \int d\Pi_n \, \sum \lvert \mathcal{M} \rvert^2 \, \frac{\vec{p}_{\chi^\dagger}{}^2}{m_\chi E_{\chi^\dagger}} e^{-E_{{\chi^\dagger}} \Delta_\chi/m_\chi} \,,
\label{eq:defS2}\end{equation}
where the normalisation constant $\mathcal{U}^S_\chi$ is chosen to ensure $\mathcal{S}_{\mathcal{T}} (T, T) = 1$, and is the expected value of $\gamma_{\chi^\dagger} v_{\chi^\dagger}$ in the collision centre of mass frame.  Using these we can write
\begin{align}
	\frac{m_\chi}{s} \, C_2^{SA} & \approx - Y_\chi \, \langle \sigma v (\chi\chi \to \chi^\dagger \phi) \rangle_{2,neq} + Y_\chi^{eq} \, \mathcal{S} (T, T_\chi) \, \langle \sigma v (\chi\chi \to \chi^\dagger \phi) \rangle_{2,eq} \notag \\
	& \quad + \frac{1}{6} \, \mathcal{U}_{\chi}^S \, x_\chi \Bigl( Y_\chi \, \langle \sigma v (\chi\chi\to\chi^\dagger\phi) \rangle_{neq} - Y_\chi^{eq} \, \mathcal{S}_{\mathcal{T}} (T, T_\chi) \, \langle \sigma v (\chi\chi\to\chi^\dagger\phi) \rangle_{eq} \Bigr) \,. \label{eq:C2SA1_simp}
\end{align}
We have introduced $x_\chi = m_\chi/T_\chi$.  See \appref{app:fullBE} for details, and more general expressions.  Note that the terms in the second line involve the usual thermally averaged cross sections of \modeqref{eq:svneq}, not those of \modeqref{eq:AnnytTA}.

When dark partners are included, we must consider the possibility that $\Psi$ will also have a temperature that deviates from the SM.  Since $\Psi$ is normally charged under the unbroken SM gauge group, it will have unsuppressed scattering with the SM plasma that remain effective till after $\chi$ has decoupled.  However, the unstable nature of $\Psi$ implies that the relevant question is not whether $\Psi$-SM scattering dominates over expansion, but whether the dark partners thermalise before decaying.  This will be true if
\begin{equation}
	n_{sc} \, \langle \sigma v \rangle_{sc} \, \frac{T}{m_\Psi} > \Gamma_\Psi \,, \qquad\qquad \text{or} \qquad\qquad  g_{sc} \, x_\Psi^{-4} \, \bigl( m_\Psi^2 \langle \sigma v \rangle_{sc} \bigr) > \frac{\Gamma_\Psi}{m_\Psi} \,, \label{eq:DPtherm}
\end{equation}
where $n_{sc}$ is the number density of the species that $\Psi$ scatters with, $g_{sc} = n_{sc}/T^3$, the factor of $T/m$ represents the typical fractional momentum transfer in collisions, and $x_\Psi = m_\Psi/T$.  We see that for this to be true out to even moderate values of $x_\Psi \sim 10^3$, we need a highly suppressed decay width, of order $\Gamma_\Psi < 10^{-15} \, m_\Psi$ for electroweak scattering.  However, this will be precisely the case we consider later.  When \modeqref{eq:DPtherm} is true, then we can take $T_\Psi = T$, but we must still include $Y_\Psi$ as a separate integration parameter; we give the relevant formulae below.  Alternatively, if the decay rate is large and \modeqref{eq:LargePsiDecay} holds, we can treat the decay as always prompt and consider the SA followed by decay as a single process.  We need only study the evolution of $Y_\chi$ and $y_\chi$.  For the former, we use the results below but with the approximation of \modeqref{eq:YPsiAppr2} to replace for $Y_\Psi$ wherever it appears; for the latter, we use \modeqref{eq:C2SA1_simp} with 
\begin{equation}
	\mathcal{S}_{\mathcal{T}} (T, T_\chi) = \frac{n_\chi^{eq} (T)}{n_\chi^{eq} (T_\chi)} \, \int dE_{\chi} \, \frac{dN_{\chi}}{dE_{\chi}} \, \frac{\vec{p}_{\chi}{}^2}{m_\chi E_{\chi}} e^{-E_{{\chi}} \Delta_\chi/m_\chi} \,,
\label{eq:modST}\end{equation}
where $dN/dE$ is the spectrum of DM produced in $\Psi$ decay, including a boost corresponding to the dark partner's momentum when produced in SA.  Finally, for intermediate widths, the dark matter temperature must be included as an additional parameter that varies with the cosmological evolution and has its own Boltzmann equation.  We defer this case to future work.

The contribution of SA and decay to the Boltzmann equations for $Y_{\chi,\Psi}$ are then simple extensions of the appropriate terms in \modeqsref{eq:dYPsi} and~\eqref{eq:dYchiAppr} to include the effects of $T_\chi$.  For SA, this is very easy since $T_\Psi = T$, and so
\begin{equation}
	\frac{dY_\chi}{dx} \supset - \frac{s}{xHZ} \, \biggl( Y_\chi^2  \, \langle \sigma v (\chi\chi \to \Psi^\dagger + SM) \rangle_{neq} - Y_\Psi \, \frac{(Y_\chi^{eq})^2}{Y_\Psi^{eq}} \, \langle \sigma v (\chi\chi \to \Psi^\dagger + SM) \rangle_{eq} \biggr) \,,
\label{eq:dYSA}\end{equation}
with the contribution to the $Y_\Psi$ evolution equation being $-2$ times this.  The complications due to DM kinetic decoupling appear in the contribution of the decay operator, which in the usual Maxwellian and non-relativistic limit is given by 
\begin{equation}
	\frac{dY_\chi}{dx} \supset \frac{\Gamma_\Psi}{x H Z} \biggl( Y_\Psi - Y_\chi \, \frac{Y_\Psi^{eq}}{Y_\chi^{eq}} \, \mathcal{D} (T, T_\chi) \biggr) \,,
\label{eq:dYDPdec}\end{equation}
where $\mathcal{D}$ is analogous to \modeqref{eq:defSTT} except for the decay operator, see \appref{app:fullBE} and in particular \modeqref{def:D} for details.  The contribution to $dY_\Psi/dx$ is exactly equal and opposite.  The SA effect on the evolution of $T_\chi$ is also a natural extension of \modeqref{eq:CollTAnn}: 
\begin{equation}
	\frac{m_\chi}{s} \, C_2^{SA} = - Y_\chi \, \langle \sigma v (\chi\chi \to \Psi^\dagger + SM) \rangle_{2,neq} + \frac{Y_\Psi}{Y_\chi} \, \frac{(Y_\chi^{eq})^2}{Y_\Psi^{eq}} \, \langle \sigma v (\chi\chi \to \Psi^\dagger + SM) \rangle_{2,eq} \,.
\label{eq:C2SADP}\end{equation}
Finally, the contribution of $\Psi$ decay is given by
\begin{equation}
	C_2^{dec} =  \mathcal{U}^D_\chi \, \frac{\Gamma_\Psi}{3 T_\chi} \biggl( \frac{Y_\Psi}{Y_\chi} - \frac{Y_\Psi^{eq}}{Y_\chi^{eq}} \, \mathcal{D}_{\mathcal{T}} (T, T_\chi) \biggr) \,,
\label{eq:C2decDP}\end{equation}
where $\mathcal{U}^D_\chi$ is the decay equivalent of $\mathcal{U}^S_\chi$, the mean of $p_\chi^2/(E_\chi m_\chi)$ produced in $\Psi$ decay; and $\mathcal{D}_{\mathcal{T}}$ is analogous to $\mathcal{S}_{\mathcal{T}}$, an integral over the decay matrix element including an exponential factor depending on $\Delta_\chi$.  Precise definitions are given in \modeqsref{eq:defgvchi} and~\eqref{def:DT} respectively in \appref{app:fullBE}.

\subsection{Asymptotic dark matter temperature}\label{subsec:TEasymp}

In \refcite{1707.09238} it was observed that SA causes a \emph{self-heating} effect that results in a different asymptotic temperature behaviour compared to ordinary cold DM.  Specifically, once DM-SM scattering decouples, ordinary DM redshifts as matter with $T_\chi \sim p_\chi \sim a^{-2} \sim T^2$, where $a$ is the expansion scale factor.  This is the reason for defining $y_\chi$ as in \modeqref{eq:defTchi}: after kinetic decoupling, $y_\chi \sim T_\chi/T^2$ is constant.  In contrast, SA converts DM mass into kinetic energy, resulting in $T_\chi \sim a^{-1} \sim T$.  Note, however, that although this is the redshift behaviour of radiation, the dark matter remains highly non-relativistic, $T_\chi \ll m_\chi$.

To understand the origin of this behaviour, let us first consider the general case.  The interaction rate for $2 \to n$ SA is $n_\chi^2 \,\langle \sigma v (SA) \rangle_{neq}$, and each such event will convert an average fraction $q_\chi$ of the mass to kinetic energy.  This fraction is process-dependent, but importantly a temperature-\emph{independent} constant.  The temperature is parametrically $T_\chi \sim E_\chi/N_\chi$, where $N_\chi$ is the total number of DM particles, and so
\begin{equation*}
	\frac{dT_\chi}{dt} \sim \frac{1}{N_\chi} \int dV \, q_\chi \, m_\chi \, n_\chi^2 \, \langle \sigma v (SA) \rangle_{neq} \sim q_\chi \, m_\chi \, n_\chi \, \langle \sigma v (SA) \rangle_{neq} \,.
\end{equation*}
Converting this to dimensionless form (and neglecting entropy production) gives
\begin{equation}
	\frac{dy_\chi}{dx} = \frac{y_\chi}{T_\chi} \, \frac{1}{xHZ} \, \frac{dT_\chi}{dt} = \frac{s^{1/3}}{xHZ} \, q_\chi \, m_\chi^2 \, Y_\chi \, \langle \sigma v (SA) \rangle_{neq} \,.
\end{equation}
The right hand side has no explicit dependence on $T_\chi$, and indeed is asymptotically constant: the cross section at late times is given by the $s$-wave piece, $Y_\chi$ approaches the constant relic value, and during radiation-domination the combination $s^{1/3}/(x H)$ is also constant.
Unless SA has no $s$-wave piece, at late times $y_\chi \propto x$ or, equivalently, $T_\chi \propto T$:
\begin{equation}
	\lim_{x\to\infty} T_\chi \approx 18 \, q_\chi \, T \, \biggl( \frac{g_{\ast S}}{80} \biggr)^{1/2} \biggl( \frac{\Omega_\chi h^2}{0.12} \biggr) \biggl( \frac{\langle \sigma v (\chi\chi \to \chi^\dagger SM) \rangle_{T = 0}}{3 \times 10^{-26} \, \text{cm}^3 \, \text{s}^{-1}} \biggr) \,.
\label{eq:dydxSAasymp}\end{equation}
We will shortly see that in practice, $q_\chi$ is at most a few percent in $2\to 2$ SA, so typically $T_\chi \approx T$, depending on the relative importance of SA versus annihilation in determining the relic density.  When the low-temperature SA cross section is enhanced, we can expect the DM to be much hotter.  For $2 \to n$ SA with $n \geq 3$, $q_\chi$ will be smaller resulting in lower temperatures, but the radiation-like redshift behaviour will remain.

The derivation of \modeqref{eq:dydxSAasymp} was based on general arguments, but it is also instructive to connect it to the results presented above.  In the absence of dark partners and neglecting terms that are exponentially suppressed at late times, the contribution of SA is
\begin{align}
	\frac{dy_\chi}{dx} & \supset \frac{s}{xHZ} \, y_\chi \, Y_\chi \, \biggl[ \frac{1}{2} \, \biggl( 1 + \frac{1}{3} \, x_\chi \, \mathcal{U}^S_\chi \biggr) \langle \sigma v (\chi\chi \to \chi^\dagger SM) \rangle_{neq} - \langle \sigma v (\chi\chi\to\chi^\dagger SM)\rangle_{2,neq} \biggr] \notag \\
	& \approx \frac{1}{6} \, \frac{s^{1/3} m_\chi^2}{xHZ} \, Y_\chi \, \mathcal{U}^S_\chi \, \langle \sigma v (\chi\chi \to \chi^\dagger SM) \rangle_{neq} \,. \label{eq:SAasymp1}
\end{align}
In the second line, we focused on the most enhanced term at low temperatures.  We see that \modeqref{eq:SAasymp1} has the same form as \modeqref{eq:dydxSAasymp}, with $q_\chi \to \mathcal{U}^S_\chi/6$.  For $2 \to 2$ SA with a massless visible final state, $\mathcal{U}^S_\chi = 9/20$ and so $q_\chi = 7.5\%$.  For dark partner models, if they decay before thermalising the above holds with the modification that $\mathcal{U}^S_\chi$ is defined from \modeqref{eq:modST}.  As discussed above, this is always true at sufficiently late times and hence gives the asymptotic behaviour.  However, even if $\Psi$-SM scattering remains efficient, we can still have this behaviour thanks to the $\Psi$ decay still converting mass to kinetic energy.  In particular, combining \modeqref{eq:C2decDP} with the late-time approximation~\eqref{eq:YPsiInf} for $Y_\Psi$ gives
\begin{equation}
	\frac{dy_\chi}{dx} \supset \frac{m_\chi \, y_\chi}{x H Z} \, \mathcal{U}^D_\chi \, \frac{\Gamma_\Psi}{3 T_\chi} \frac{1}{Y_\chi} \, \biggl( \frac{s \, \langle \sigma v (\chi \chi \to \Psi^\dagger V) \rangle_{T=0}}{2 \Gamma_\Psi} \, Y_\chi^2 \biggr) = \frac{1}{6} \, \frac{s}{x H Z} \, y_\chi \, Y_\chi\, x_\chi \, \mathcal{U}^D_\chi \, \langle \sigma v\rangle_{T=0} \,,
\end{equation}
which again reproduces \modeqref{eq:dydxSAasymp} with $q_\chi \to \mathcal{U}^D_\chi/6$.

%% file: Files/BW.tex
\section{Late-time resonantly enhanced semi-annihilation}\label{sec:bw}

In the previous section we outlined how the presence of SA modifies the thermal evolution of dark matter.  Most important is the general feature of \modeqref{eq:dydxSAasymp}, that at late times $T_\chi$ is proportional to, and potentially larger than, the SM temperature.  However, in most models we expect kinetic decoupling to occur \emph{after} chemical decoupling, making this feature less important.  One possible signal was discussed in \refcite{1805.05648}, where for light DM the different kinetic behaviour can modify small scale structure.  Here we instead consider scenarios where the cross section is modified at low temperatures, such that number-changing processes continue to be important during epochs when $T_\chi \neq T$.

Specifically we consider weak- or TeV-scale DM where the (semi)-annihilation cross section features a narrow $s$-channel resonance, which we can model as a Breit-Wigner peak,
\begin{equation}
	\sigma v \propto \frac{1}{(E^2 - M^2)^2 + M^2 \Gamma^2} \,.
\label{eq:BWbasic}\end{equation}
Here, $E$ is the centre of mass energy, and $M$ and $\Gamma$ are the mass and width of the intermediate resonance.  At relatively high temperatures $x \sim 25$, the thermal kinetic energy is large compared to the width, $E - M \gg \Gamma$, and the first term in the numerator dominates.  As the temperature drops, $E - M$ becomes comparable to the width and the total rate is enhanced. Such models have been considered as possible ways to enhance signals in indirect detection experiments while retaining thermal freeze-out~\cite{0812.0072,0901.1450,0903.0122,1706.09974,1705.10777,1707.09313}, and in particular as ways to explain the positron excess seen by AMS-02 and earlier experiments.

Na\"\i vely one can obtain arbitrarily large cross sections by taking the width sufficiently small, at the cost of fine-tuning the model parameters.  However, it was shown in \refcite{1106.6027} that the possible enhancement available through this mechanism is bounded from above by $\order{10^2}$ for TeV DM.  This is due to a period of enhanced annihilation during kinetic decoupling, when the DM temperature decreases rapidly and the typical kinetic energy becomes closer to the resonance.  To see this more clearly, let us briefly review their argument.  We define $E^2 = 4m^2 (1 + z)$, $M = 2m \sqrt{1- \delta}$, and $M \Gamma = 4m^2 \gamma$, where $m$ is the mass of the initial state particles; then the cross section \eqref{eq:BWbasic} takes the form
\begin{equation}
	\sigma v = \sigma_0 \, \frac{\delta^2 + \gamma^2}{(\delta + z)^2 + \gamma^2} \,,
\label{eq:BWzdg}\end{equation}
where $\sigma_0 = \sigma v \rvert_{z=0}$.  For $\delta > 0$ and in the non-relativistic limit, this has thermal average~\cite{0812.0072}
\begin{equation}
	\expect{\sigma v} \approx \sigma_0 \, \frac{\delta^2 + \gamma^2}{(\delta + \xi x^{-1})^2 + \gamma^2} \,,
\label{eq:avBW}\end{equation}
where $\xi \approx 1/\sqrt{2}$ is a constant.  Note that this will continue to increase with temperature till $x = x_b \approx \max [\delta, \gamma]^{-1}$.  A simple estimate for the relic density is to integrate the Boltzmann equation only over the epoch where both the equilibrium density can be neglected and the cross section is constant.  This tells us that the correct relic abundance is obtained for 
\begin{equation}
	\frac{\sigma_0}{\sigma_c} \approx \frac{x_b}{x_f} \approx \frac{\max [\delta, \gamma]^{-1}}{25} \,,
\label{eq:simplenokd}\end{equation}
where $\sigma_c \approx 3 \times 10^{-26}$\,cm$^3$\,s$^{-1}$ is the usual thermal cross section.  \Modeqref{eq:simplenokd} assumes equality of the temperatures of the two sectors; if the dark and visible sectors kinetically decouple at some temperature $x_{kd} \in (x_f, x_b)$, then we must replace $x_b$ with the SM temperature when $T_\chi = m_\chi/x_b$.  For conventional DM that redshifts as matter, $x_\chi = x^2/x_{kd}$, this temperature is $x_b' \approx \sqrt{x_{kd}} \max [\delta, \gamma]^{-1/2} < x_b$, resulting in a smaller value of $\sigma_0$.  To get the maximum possible cross section today we must evaluate \modeqref{eq:BWzdg} for $z \sim 10^{-6}$,
\begin{equation}
	\sigma_G \approx \sigma_0 \, \frac{\max [\delta, \gamma]^2}{\max [\gamma^2, \delta^2, 2 \delta z]} \sim \sigma_0 \min \bigl[ 1, 10^6 \max [\delta, \gamma] \bigr] \,.
\label{eq:actualBWE}\end{equation}
This suggests that the maximum possible boost occurs for $\delta \sim \gamma \sim 10^{-6}$, which is supported by more careful numerical calculations of \refcite{1106.6027}.  Finally, that work made the additional assumption that thermal contact between the SM and DM is maintained by the same couplings responsible for annihilation.  That implies the upper limit $\sigma_G \lesssim 10^2 \sigma_c$; and the maximum enhancement is only obtained when $x_{kd}$ is comparable to the usual temperature of freeze-out, $x_f \sim 25$.

The last assumption is key.  It is well-motivated both on the grounds of minimality, and that any coupling which allows the DM to scatter off the SM will also contribute to DM annihilation.  But there are two obvious loopholes, both of which we will exploit.  First, for any DM model, interactions that lead to $p$-wave annihilation will be suppressed compared to $s$-wave processes that determine the relic density and indirect detection signals, but can still dominate the scattering rate.  This requires multiple mediators between the dark and visible sectors, but this is not rare in top-down models and can even be required for UV-completeness~\cite{1510.02110,1612.03475,1610.03063,1611.04593}.  In models where SA (dominantly) determines the relic density, any scattering \emph{must} involve different (combinations of) interactions, and so they naturally fall into this category.

The second loophole is precisely the point discussed in \secref{subsec:TEasymp}: after kinetic decoupling, the temperature of SADM \emph{does not} redshift as matter.  Additionally, for cross sections which are enhanced at low temperatures, \modeqref{eq:dydxSAasymp} demonstrates that the temperature after decoupling is typically higher than the SM.  This could potentially allow $x_b' > x_b$, increasing the enhancement over what would be expected from \modeqref{eq:simplenokd}.  It also raises the possibility of models where scattering is absent, and the DM temperature is maintained only by SA; however, we defer this more exotic possibility to future work.

There is one additional extension we make.  It was shown in \refcite{1705.10777} that \modeqref{eq:BWbasic} is an inadequate approximation when the resonance is nearly on-shell.  The Breit-Wigner propagator is an approximation that neglects some of the energy dependence.  Specifically, the Green's function is
\begin{equation}
	D_F^{-1} (E) = E^2 - M_0^2 - \mathcal{M}_2^2 (E) = E^2 - M_0^2 - \Re \mathcal{M}_2 (E) - i \, \Im \mathcal{M}_2 (E) \,,
\end{equation}
where $M_0$ is the bare (unrenormalised) mass and $\mathcal{M}_2$ encodes loop corrections.  Ordinarily it is satisfactory to take the loop correction as a constant evaluated at the physical mass $M$, where the Cutkosky rules give $\Im \mathcal{M}_2 (M) = M \, \Gamma$.  However, this breaks down in the presence of a threshold for a new decay, where the imaginary part changes rapidly.  Small $\delta$ by definition means that non-relativistic processes are taking place near such a threshold, and is especially prominent from taking $z \sim \delta$ in \modeqref{eq:actualBWE}.  To correct for this effect we make the replacement $M \, \Gamma \to E \, \Gamma(E)$, where $\Gamma (E)$ is the width the resonance \emph{would} have, if it had mass $E$ (and all other model parameters where unchanged).  We also make the natural definition $4m^2 \, \tilde{\gamma} (z) = E \, \Gamma (E)$.  For $\delta > 0$, the on-shell resonance cannot decay to DM, but that process \emph{will} contribute to $\Gamma (E)$ for all $E \geq 2m$.  Large enhancements require small resonance-DM couplings, so that $\Gamma (E) \sim \Gamma$ near threshold.  Since we need $\Gamma(E) \sim 10^{-6} M$ to maximise \modeqref{eq:actualBWE}, the \emph{actual} width that maximises the enhancement will be slightly smaller.  However, since $\Gamma (E)$ is a monotonically increasing function, the width at low temperatures is less than at high, and so this represents a modest increase in the overall enhancement.

%% file: Files/model.tex
\section{UV-complete simplified model}\label{sec:model}

We now construct a simplified model to serve as a case study of the phenomenology discussed so far.  Our intention is to fit the AMS-02 positron excess using the process $\chi\chi \to \Psi^\dagger \mu_R$, so our theory must contain a minimum of three states charged under the dark sector symmetry: the dark matter $\chi$, a fermion dark partner $\Psi$, and a bosonic resonance $\Phi$.  We take the minimal case for $\Phi$ (a SM-singlet complex scalar) and $\Psi$ (a Dirac fermion).  We consider fermion dark matter to avoid a tree-level Higgs portal annihilation channel.  As discussed in \secref{sec:DP}, the symmetry that stabilises $\chi$ must be equivalent to a global $\set{Z}_3$, so for simplicity we make that choice.

A model with this dark matter and dark partner was among those catalogued using effective field theories in \refcite{1611.09360}, which also discussed the possible choices for the dark partner decay.  The minimal possibility involving muon final states is the three-body decay $\Psi \to \chi \mu\bar{\nu}$, which can be derived from the dimension-7 operator
\begin{equation}
	\lag_{decay} = \frac{1}{\Lambda^3} \, (\bar{\Psi} \mu_R) \, \bigl( (\bar{L} \tilde{H}) \chi \bigr) \,,
\label{eq:dpdec}\end{equation}
where $\tilde{H} = i \sigma^2 H^\ast$ and $L$ is the (left-handed) lepton doublet.  Constructing a UV-complete theory requires opening this operator into renormalisable terms.  There are eight possible ways to do this at tree-level.  We choose the unique option which introduces only one additional degree of freedom: a complex scalar doublet $\Sigma$, charged under $\dsym$ and coupling $\chi$ and $L$.  This Yukawa coupling will also lead to $\chi L \to \chi L$ scattering that maintains thermal contact to $x \sim 10^3$, provided that $\Sigma$ is not too heavy.

\begin{table}
	\centering
	\begin{tabular}{|c|c|c|c|}
		\hline
		Particle & Spin & $\dsym = \set{Z}_3$ & $SU(2)_L \times U(1)_Y$ \\
		\hline
		$\chi$ & $1/2$ & $\exp{2\pi i/3}$ & $1_0$ \\
		$\Psi$ & $1/2$ & $\exp{2\pi i/3}$ & $1_{-1}$ \\
		$\Phi$ & 0 & $\exp{2\pi i/3}$ & $1_0$ \\
		$\Sigma$ & 0 & $\exp{2\pi i/3}$ & $2_{1/2}$ \\
		\hline
	\end{tabular}
	\caption{New particle content in our simplified model}\label{tab:npf}
\end{table}

The sum total of new particles is listed in \tabref{tab:npf}.  The Lagrangian is
\begin{equation}
	\lag = \lag_{SM/V(H)} + \lag_{kin} + \lag_Y - V(H, \Phi, \Sigma) \,,
\end{equation}
with kinetic terms
\begin{equation}
	\lag_{kin} = \bar{\chi} \bigl(i \partial\!\!\!/ - m_\chi) \chi + \bar{\Psi} (i D\!\!\!\!/ - m_\Psi) \Psi + \abs{\partial_\mu \Phi}^2 + \abs{D_\mu \Sigma}^2 ,
\end{equation}
Yukawa interactions
\begin{equation}
	\lag_Y = g_\chi \, \Phi \, \bar{\chi}^c \gamma^5 \chi + g_\Psi \, \Phi \, \bar{\Psi} \mu_R + g_\Sigma \, \Sigma \, \bar{\chi} L + h.c. \,,
\label{eq:lagyuk}\end{equation}
and scalar potential
\begin{align}
	V(H, \Phi, \Sigma) & = \frac{1}{2} \, \lambda_h \biggl(H^\dagger H - \frac{1}{2} v_h^2 \biggr)^2 + m_\Phi^2 \, \Phi^\dagger \Phi + m_\Sigma^2 \, \Sigma^\dagger \Sigma + \frac{1}{2} \, \lambda_\Phi (\Phi^\dagger \Phi)^2 + \frac{1}{2} \, \lambda_\Sigma (\Sigma^\dagger \Sigma)^2 \notag \\
	& \quad + \lambda_{h\Phi} \, \biggl(H^\dagger H - \frac{1}{2} v_h^2 \biggr) \, \Phi^\dagger \Phi + \lambda_{h\Sigma} \, \biggl(H^\dagger H - \frac{1}{2} v_h^2 \biggr) \, \Sigma^\dagger \Sigma + \lambda_{\Phi\Sigma} \, \Phi^\dagger \Phi \, \Sigma^\dagger \Sigma \notag \\
	& \quad + \biggl( \mu_\Sigma \, \Phi \, \Sigma^\dagger H + \frac{1}{6} \, \mu_\Phi \, e^{i \alpha_\Phi} \Phi^3 + \frac{1}{2} \, \lambda_3 \, e^{i \alpha 3} \, (\Phi^\dagger)^2 \Sigma^\dagger H + h.c. \biggr)\,. \label{eq:scapot}
\end{align}
We have used phase rotations to set the Yukawa couplings and $\mu_\Sigma$ to be real and positive without loss of generality.  All other couplings are automatically real except the last two on the last line of \modeqref{eq:scapot}, where we have written the phases explicitly.  We chose the $\Phi$-dark matter Yukawa to be a pseudoscalar-like coupling in order to have $s$-wave semi-annihilation.

Note that if we did not include the $\Sigma$ field, there would be an additional accidental $\set{Z}_4$ symmetry under which $\chi \to i \chi$, $\Psi \to - \Psi$ and $\Phi \to - \Phi$.  This would enforce dark partner stability for the mass range of interest, $m_\Psi < 2 m_\chi, m_\Phi$.  Further, this symmetry is restored in the limit $\mu_\Sigma, \mu_\Phi, \lambda_3 \to 0$, which means it is technically natural for all three couplings to be simultaneously small.  In \appref{app:UV}, we outline a theory where $\dsym$ is a residual subgroup of $U(1)_{L_\mu - L_\tau}$, where $\mu_\Phi = \lambda_3 = 0$ and $\mu_\Sigma$ is technically natural by itself.  This alternate theory can also explain the particular flavour structure of the Yukawa couplings, by forbidding interactions with the tau or electron prior to the breaking of the flavour symmetry.

The scalar cubic coupling $\mu_\Sigma$ introduces a mass mixing between the resonance $\Phi$ and the neutral component $\Sigma^0$ of the extra doublet.  This mixing is necessary to allow the dark partner to decay, but the coupling introduces an additional decay mode for the resonance, $\Phi \to \Sigma H^\dagger$.  This must be subdominant to the decay to the dark partner, since it will produce a large number of soft photons from Higgs and gauge boson decay and is subject to stringent constraints from CMB observations and galactic centre observations.  For $\Gamma_\Phi \sim 10^{-6} \, m_\Phi$, the relevant region of parameter space is
\begin{equation}
	g_\Psi \lesssim 10^{-2} \,, \qquad \mu_\Sigma \lesssim 5 \times 10^{-3} \, m_\Phi \,.
\label{eq:decparam}\end{equation}
For TeV-scale DM, this corresponds to GeV-scale $\mu_\Sigma$.

We introduced the coupling $g_\Sigma$ as a UV completion of \modeqref{eq:dpdec} to allow $\Psi$ to decay.  However, it also leads to two other phenomenologically important processes: DM-SM scattering and DM annihilation to the visible sector.  In particular, we want the former to be important while keeping the latter small.  This can be achieved as the scattering process is also resonantly enhanced (though much less so than our main processes of interest).  We can estimate the kinetic decoupling temperature $T_{kd}$ by equating the expansion rate to the scattering momentum transfer rate,
\begin{equation}
	H (T_{kd}) \approx n_L \, \langle \sigma v \rangle_s \, \frac{T_{kd}}{m_\chi} \,,
\end{equation}
where $n_L$ is the number density of $L$, and the factor of $T_{kd}/m_\chi$ represents the approximate momentum transfer per collision.  This gives us a relation between $y_\Sigma$ and $x_{kd} = m_\chi/T_{kd}$, which we can then insert in the annihilation cross section:
\begin{equation}
	\sigma_{an} v = 0.77\; \text{pb} \, \biggl( \frac{x_{kd}}{10^3} \biggr)^4 \biggl( \frac{\text{1 TeV}}{m_\chi} \biggr)^2 \biggl( \frac{m_\Sigma^2 - m_\chi^2}{m_\Sigma^2 + m_\chi^2} \biggr)^2 \,.
\end{equation}
Compared to the thermal relic cross section, this is already suppressed by over an order of magnitude for $m_\Sigma \approx 1.4\,m_\chi$, with greater suppression for lighter $\Sigma$.  Annihilation will be a subleading correction to the thermal relic density, and negligible compared to SA for indirect detection today.

There are two important qualitatively different phases in this model, according to whether the dark partner decays through the direct three-body (3B, $\Psi \to \chi\mu\nu$) or sequential two-body (S2B, $\Psi \to \Sigma \mu$, $\Sigma \to \chi \nu$) channel.  This will lead to different spectra of the muon decay product, and ultimately of the positrons produced in muon decay.  In particular, the latter process produces a harder spectrum, where the muon is monochromatic in the dark partner rest frame.  In the following section, we will see how this qualitative difference affects the ability to explain the positron excess.

%% file: Files/amsfit.tex
\section{Fitting the AMS-02 positron excess}\label{sec:fit}

In this section we investigate how well the DM model described above can explain the AMS-02 positron excess.  We do not consider factors such as the relic density or other experimental constraints, but simply focus on what masses and cross sections provide the optimal fit to the data.  Rather than attempt a full parameter scan, we consider two ansatzes for the ratios of dark sector masses $\rho_\Psi \equiv \frac{m_\Psi}{m_\chi}$ and $\rho_{\Sigma}\equiv \frac{m_\Sigma}{m_\chi}$, namely
\begin{equation}
	\{\rho_{\Sigma}, \rho_\Psi \} = 
	\begin{cases}
		\{ 1.2, 1.3\} & \text{S2B,} \\
		\{ 1.4, 1.3\} & \text{3B.}
	\end{cases}
\end{equation} 
This reduces the relevant parameter space to the DM mass $m_\chi$ and SA cross section $\langle \sigma v (\chi\chi \to \bar{\Psi}\mu)\rangle$.  Other processes can be neglected as they are either not resonantly enhanced (\emph{e.g.} annihilation $\chi\bar{\chi} \to \mu\bar{\mu}$), or suppressed by small $\mu_\Sigma$ as discussed around \modeqref{eq:decparam}.

\begin{figure}
	\centering
	\includegraphics[width=0.5\textwidth]{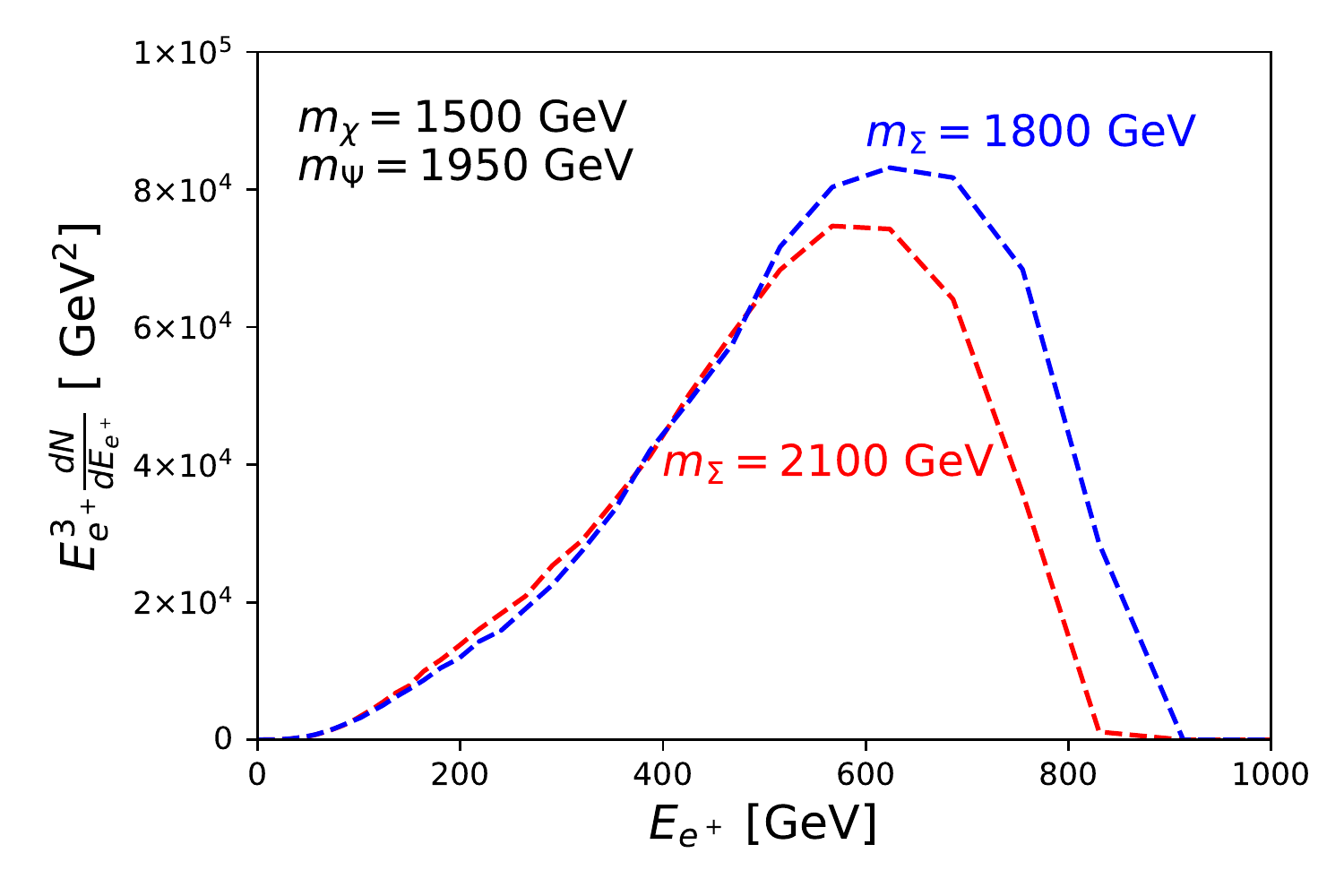}
	\caption{Positron spectrum at production for the two dark partner decays modes, with S2B (3B) in blue (red).  The specific parameters are as labelled.}\label{fig:origspec}
\end{figure}

To study DM (semi)-annihilation, we implement this model in \texttt{FeynRules}~\cite{1310.1921}.  Since DM in the galaxy today has a velocity $v \sim 10^{-6}$, we can efficiently simulate the kinematics of the final states using only the decay of the resonance $\Phi$.  We generate parton-level events for the resonance decay in \texttt{MadGraph}~\cite{1405.0301}.  Showering and hadronisation are later performed with \texttt{PYTHIA 8.2}~\cite{1410.3012}.  We cross-checked the positron spectra per annihilation derived this way by using the same tools to analyse DM annihilation to muon pairs, and comparing with \texttt{PPPC 4 DM ID}~\cite{1009.0224}.  There are minor discrepancies for positron energies above a few hundred GeV, due to the absence of electroweak (EW) corrections in our spectra.  However, experimental uncertainties at high positron energies are relatively large, such that including EW corrections would hardly change the results of the analysis.  We show the positron spectrum generated this way in \figref{fig:origspec} for our two choices of $\rho_{\Psi,\Sigma}$ and a fixed DM mass $m_\chi = 1.5$\,TeV.  The two spectra are very similar, though we see that the S2B decay mode leads to slightly harder positrons, as expected based on our qualitative arguments of the previous section. 

\begin{table}
	\centering
	\begin{tabular}{ccccc}
		\hline
		Model & $\delta$ & $K_0\,[{\rm kpc}^2/{\rm Myr}]$   & $L\,[{\rm kpc}]$ & $V_c\,[{\rm km}/{\rm s}]$ \\
		\hline
		MIN   & 0.85     & 0.0016  &  1  &  13.5 \\
		MED   & 0.70     & 0.0112  &  4  & 12    \\
		MAX   & 0.46     & 0.0765  & 15  & 5     \\
		\hline
	\end{tabular}
	\label{tab:prop}\caption{Parameters for propagation models MIN, MED and MAX.  $L$ is the height of the galactic disc, $V_c$ the magnitude of the galactic wind, and $K_0$ and $\delta$ are the parameters of \modeqref{eq:diff}.}
\end{table}

The propagation of the positron spectra from their point of origin to detection can be performed using the diffusion-loss equation~\cite{1012.4515},
\begin{equation}
	\frac{\partial f}{\partial t} - \nabla \bigl( \mathcal{K} (E, \vec{x}) \nabla f \bigr) + \vec{V}_c \cdot \vec{\nabla} f - \frac{\partial}{\partial E} \bigl( b (E, \vec{x}) f \bigr) = Q(E, \vec{x}) \,,
\label{eq:diffeq}\end{equation}
where $f(t, \vec{x}, E)$ is the positron number density, $\mathcal{K}$ the diffusion coefficient, $V_c$ the galactic wind, $b$ the energy loss coefficient function and $Q$ the source term.  This last object contains the particle physics input:
\begin{equation}
	Q = \frac{1}{2} \, n_\chi^2 \, \sum_f \langle \sigma v \rangle_f \, \frac{dN^f_{e^+}}{dE} \approx \frac{1}{4} \, n_\chi^2 \, \langle \sigma v (\chi\chi \to \bar{\Psi} \mu) \rangle \, \frac{dN^{SA}_{e^+}}{dE} \,,
\end{equation}
where we assume that SA dominates due to the resonant enhancement.  The additional factor of one-half is due to a difference in how the cross sections are normalised; $\langle \sigma v \rangle_f$ averages over both particle and anti-particle, while $\langle \sigma v (\chi\chi \to \bar{\Psi} \mu) \rangle$ does not.  The remaining terms in \modeqref{eq:diffeq} describe the propagation itself, and different choices of these parameters allow us to explore the astrophysical uncertainties.  As the goal of this work is to explain the positron spectrum with dark matter physics, we restrict ourself to the choice of several benchmark scenarios.  For the diffusion coefficient and galactic wind, we consider the MIN, MED and MAX models defined in \refscite{0712.2312, astro-ph/0306207} as illustrative of the range of possibilities.  These model the galaxy as a disc of varying height $L$ and parameterise the diffusion coefficient as
\begin{equation}
	\mathcal{K} (E) = K_0 \, (E/\text{GeV})^\delta \,,
\label{eq:diff}\end{equation}
with the specific model parameters given in \tabref{tab:prop}.  The energy loss function depends on the galactic magnetic field, for which we use the parameterisation of \refcite{1505.01049},
\begin{equation}
	B_{tot} = B_0 \, \exp \biggl( - \frac{r - r_\odot}{r_D} - \frac{\lvert z \rvert}{z_D} \biggr) \,,
\label{eq:mf}\end{equation}
where $(r, z)$ are the Galactic coordinated, $r_\odot$ the location of the Sun, and the different choices of $B_0$, $r_D$ and $z_D$ are given in \tabref{tab:mf}.  Other possible uncertainties come from the choice of DM density profile.  As detailed in the following section, cuspy profiles such as NFW or Einasto are heavily constrained by the absence of a $\gamma$-ray signal from the galactic centre, so we use an isothermal profile (\texttt{Iso}) as representative of a cored distribution.  With these choices, we can integrate the diffusion equation using the numerical Green's functions provided by \texttt{PPPC 4 DM ID}~\cite{1505.01049}.

\begin{table}
	\centering
	\begin{tabular}{cccc}
		\hline
		Magnetic Field & $B_0\,[\mu G]$ & $r_D\,[{\rm kpc}]$   & $z_D\,[{\rm kpc}]$ \\
		\hline
		MF1   & 4.78  & 10  &  2  \\
		MF2   & 5.1    & 8.5  &  1  \\
		MF3   & 9.5    & 30  & 4  \\
		\hline
	\end{tabular}
	\label{tab:mf}\caption{Parameters for the magnetic field parameterisation of \modeqref{eq:mf}.}
\end{table}

In addition to the primary positrons from SA, we must include the astrophysical contribution to the positron flux.  Pulsars are a potential source of primary positrons, and have been used as an alternative explanation for the excess~\cite{0810.1527,0903.1310,1108.4827}.  As we are focusing on DM explanations, we do not include any such contribution.  Secondary cosmic positrons are produced by spallation reactions of cosmic nuclei with interstellar medium.  The predicted spectra vary with the choice of modelings of nuclear cross sections and the propagation parameters~\cite{0809.5268}.  We adopt the secondary positron flux proposed in the model from \refcite{astro-ph/9808243} and parameterised in \refcite{0802.3378} as
\begin{equation}
	\frac{d\Phi_{e^+}^{sec}}{d{\rm E}_{e^+}}=\frac{4.5 \, {\rm E}_{e^+}^{0.7}}{1 + 650 \, {\rm E}_{e^+}^{2.3}+1500 \, {\rm E}_{e^+}^{4.2}} \; 
	{\rm GeV}^{-1} {\rm cm}^{-2} {\rm s}^{-1} {\rm sr}^{-1} \; ,
\label{eq:esec}\end{equation}
where ${\rm E}_{e^+}$ is the positron energy in GeV.

\begin{figure}
	\includegraphics[width=0.45\textwidth]{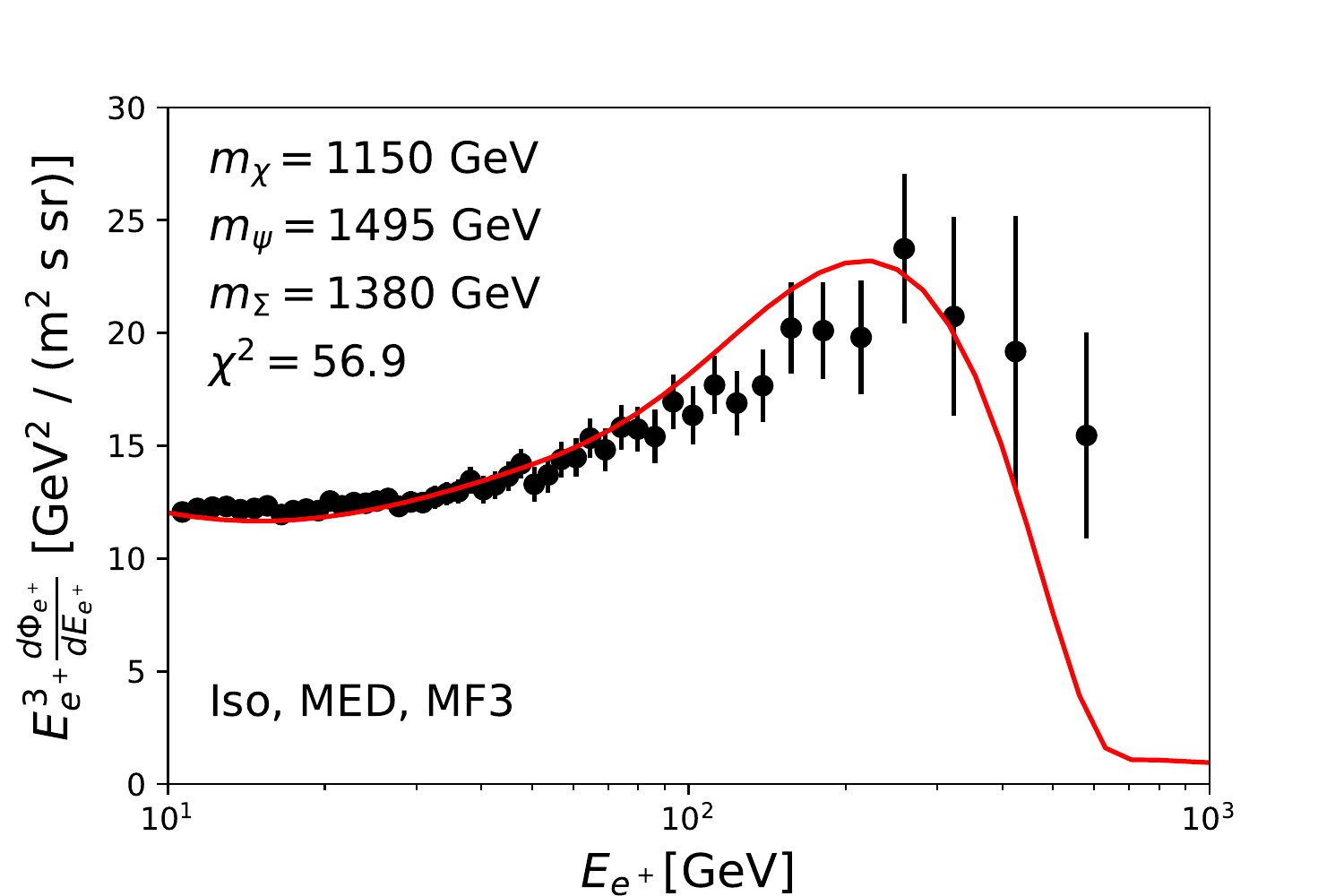}
	\hfill
	\includegraphics[width=0.45\textwidth]{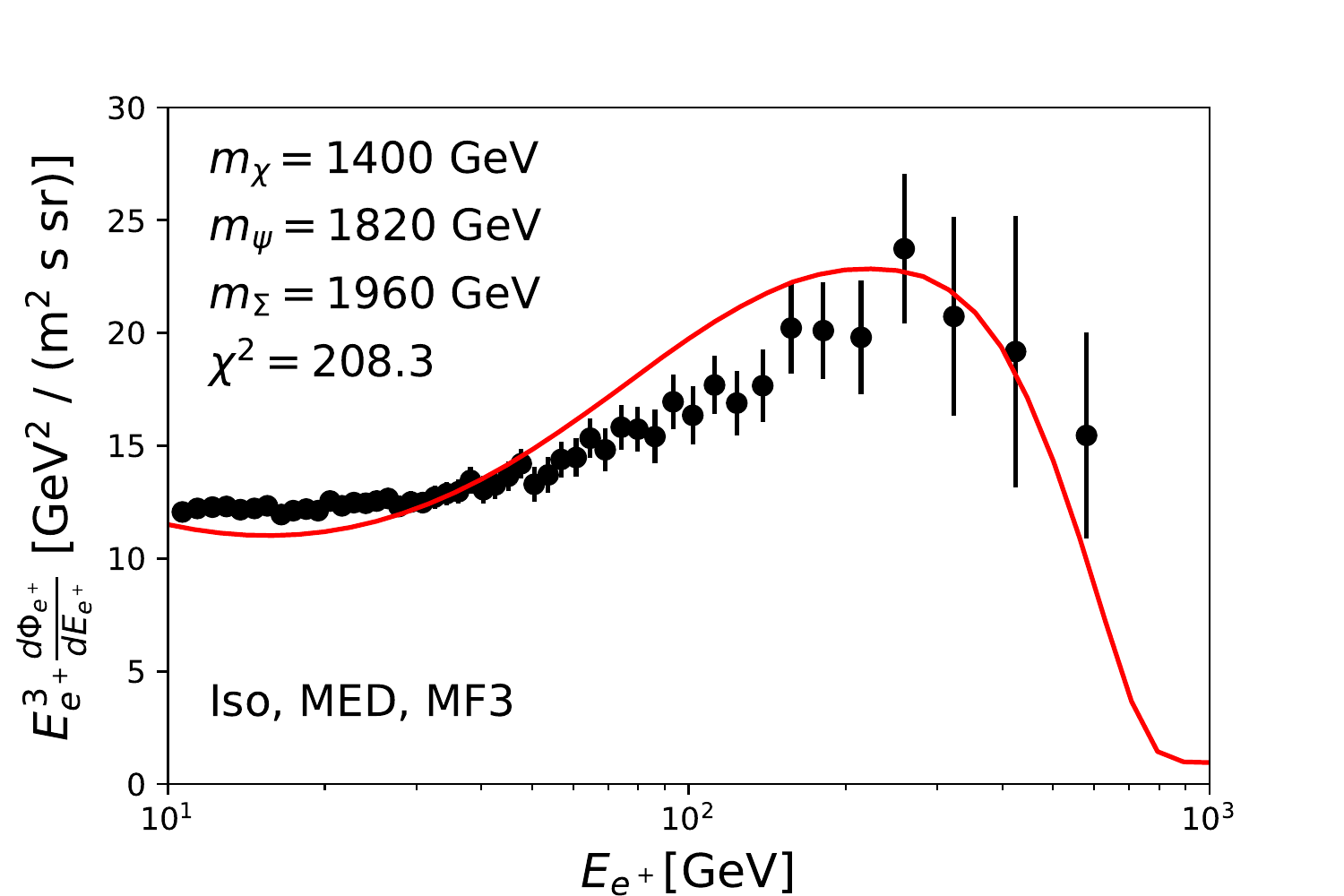}
	\caption{The best fit spectra for the S2B (left) and 3B (right) cases.  The points mark the AMS data with error bars.}\label{fig:bestfits}
\end{figure}

With these choices, we vary our parameters in the DM model space and over our discrete set of astrophysical ansatzes.  We add the positron spectra generated this way to the secondary flux of \modeqref{eq:esec}.  We restrict our attention to energies E$_{e^+} > 10$\,GeV; below this point, the contribution to the flux from DM is negligible compared to the secondary background, while the experimental uncertainties are very small, and the quality of our fit would be driven by astrophysics rather than the particle physics we focus on.  We show the best fit spectra for each decay mode in \figref{fig:bestfits}; the associated SA cross sections are $\langle \sigma v \rangle = 3.5\times 10^{-23} \,{\rm cm}^3\,{\rm s}^{-1}$ and $2.6\times 10^{-23} \,{\rm cm}^3\,{\rm s}^{-1}$ respectively, with a statistical uncertainty of 1--2\%.  For positron spectrum with energy larger than 10 GeV, the best fits have $\chi^2 = 56.9 $ and $208.3$ for 49 degrees of freedom.

%% file: Files/rdens.tex
\section{Thermal relic density and late-time cross section enhancement}\label{sec:rd}

In the previous section, we showed that SA can provide a good fit to the AMS positron excess for TeV-scale DM and a cross sections $\sigma (\chi\chi \to \Psi^\dagger \mu) \approx 3 \times 10^{-23}$\,cm$^3$\,s$^{-1}$.  As expected, this cross section is several orders of magnitude larger than the thermal relic cross section.  In this section we show that the model of \secref{sec:model} can simultaneously explain this signal cross section and provide the correct relic density due to the $s$-channel resonance.  We also discuss other experimental bounds, most notably from the CMB but also other indirect channels such as gamma rays.

The majority of the technical details for $\chi$ freeze out have already been discussed in a general context in \secsref{sec:TempEvo} and~\ref{sec:bw}, but there are some specific points in this model that need explanation.  Most important is the presence of the additional scalar $\Sigma$, which can be produced by $\chi\chi$ SA and dark partner decay, and has two-body decay to DM.  All our Boltzmann equations will explicitly depend on $f_\Sigma$, which in general requires its own set of evolution equations.  However, we exploit the relatively large width for the two-body decay, $\Gamma_\Sigma/m_\Sigma \sim g_\Sigma^2/8\pi \sim 10^{-3}$.  As discussed below \modeqref{eq:LargePsiDecay}, this decay dominates over both expansion and production prior to $\chi$ freeze-out, especially given that SA production is somewhat suppressed.  We can use a variant of \modeqref{eq:YPsiAppr2} to approximate for $Y_\Sigma$ at all times:
\begin{equation}
	Y_\Sigma \approx Y_\chi \, \frac{Y_\Sigma^{eq}}{Y_\chi^{eq}} \, \mathcal{S}^{(\Sigma)} (T, T_\chi) + \frac{1}{2} \, \Theta (m_\Psi - m_\Sigma) \, \frac{\Gamma_\Psi}{\Gamma_\Sigma} \, Y_\Psi + \frac{s \, \langle \sigma v (\chi \chi \to \Sigma^\dagger H) \rangle_{neq}}{4 \Gamma_\Sigma} \, Y_\chi^2 \,,
\label{eq:Ysig}\end{equation}
where $\Theta$ is the Heaviside-Lorentz function.  We included an additional term for the contribution of dark partner decay.  This is relevant despite the small ratio of widths, $\Gamma_\Psi/\Gamma_\Sigma \lesssim 10^{-10}$, because $Y_\Psi \propto \Gamma_\Psi^{-1}$ at late times as can be seen from \modeqref{eq:YPsiInf}.  We also corrected the contribution from inverse decay due to $T_\chi \neq T$ using a variation of \modeqref{eq:defSTT},
\begin{equation}
	\mathcal{S}^{(\Sigma)} (T, T_\chi) = \frac{n_\chi^{eq} (T)}{n_\chi^{eq} (T_\chi)} \, \int dE_{\chi} \, \frac{dN_{\chi}}{dE_{\chi}} \approx \exp \biggl( - \frac{(m_\Sigma - m_\chi)^2}{2m_\Sigma m_\chi} \, \Delta_\chi \biggr) \,,
\end{equation}
with $dN_{\chi}/dE_{\chi}$ the spectrum of DM produced in $\Sigma$ decay.  The second equality assumes the non-relativistic limit and that $\Sigma$ is at rest; this will be a good approximation if inverse decay dominates production, which is when this correction is most important.  Accordingly, we use \modeqref{eq:Ysig} to replace for $Y_\Sigma$ whenever it appears in our Boltzmann equations.  For computing the contributions of inverse processes, we assume $\Sigma$ is in kinetic equilibrium with the SM; this is true whenever they are relevant, as inverse decay processes dominate $\Sigma$ production during these early epochs.  However, when computing the effect of SA on $T_\chi$, we follow the discussion on prompt decays in \secref{subsec:TESA} and approximate production and decay as a single process.  

The dark partner lifetime is highly suppressed by our demand that $\Sigma$-$\Phi$ mixing be small.  Even for the more rapid two-body decay, the lifetime is parametrically
\begin{equation}
	\frac{\Gamma_\Psi}{m_\Psi} \sim \frac{g_\Psi^2}{32\pi} \, \sin^2 \theta \, \biggl(1 - \frac{m_\Sigma^2}{m_\Psi^2} \biggr)^2 \sim 10^{-15} \,,
\end{equation}
where $\theta$ is the appropriate mixing angle.  The three-body decay is naturally even smaller.  Note that as discuss in \secref{subsec:TESA}, this is a slow enough decay that the DP thermalises before decay till at least $x \gtrsim 10^3$.  We therefore make the approximation that $\Psi$ has the same temperature as the SM in evaluating processes that involve it.

We are left with three Boltzmann equations, for $Y_{\chi,\Psi}$ and $y_\chi$.  We numerically integrate these till $Y_\chi$ and $y_\chi/x$ are approximately constant, and $Y_\Psi$ is given by \modeqref{eq:YPsiInf}.  We cross-check our numerical code by comparing to micrOMEGAs~4.1~\cite{1407.6129,1509.08481} for semi-annihilating non-resonant models with $T_\chi = T$, from which we include a 15\% theoretical uncertainty in the relic density.  Our model contains a large number of parameters; at a minimum, the relic density depends on four masses, three Yukawas, and the cubic coupling $\mu_\Sigma$.  Rather than attempt a detailed scan of the whole model space, we focus on the values of $m_\chi$, $m_\Psi$ and $m_\Sigma$ corresponding to our two best-fit points in the previous section.  We fix the $\Sigma$-$\Phi$ mixing angle $\theta$ to discrete values, which determines the cubic coupling; set $g_\Sigma = 1$, to maximise the scattering cross section while remaining perturbative; use the relic density to fix $g_\chi$; and convert the remaining parameters $\{m_\Phi,\, g_\Psi\} \to \{\delta, \gamma\}$.  We also apply the conservative constraint that $g_\chi \leq 1$ to ensure numerical stability; as discussed in \secref{sec:bw}, we expect large enhancements to require small DM-resonance coupling, so this should not significantly effect the preferred parameter space.

\begin{figure}
	\centering
	\includegraphics[width=0.45\textwidth]{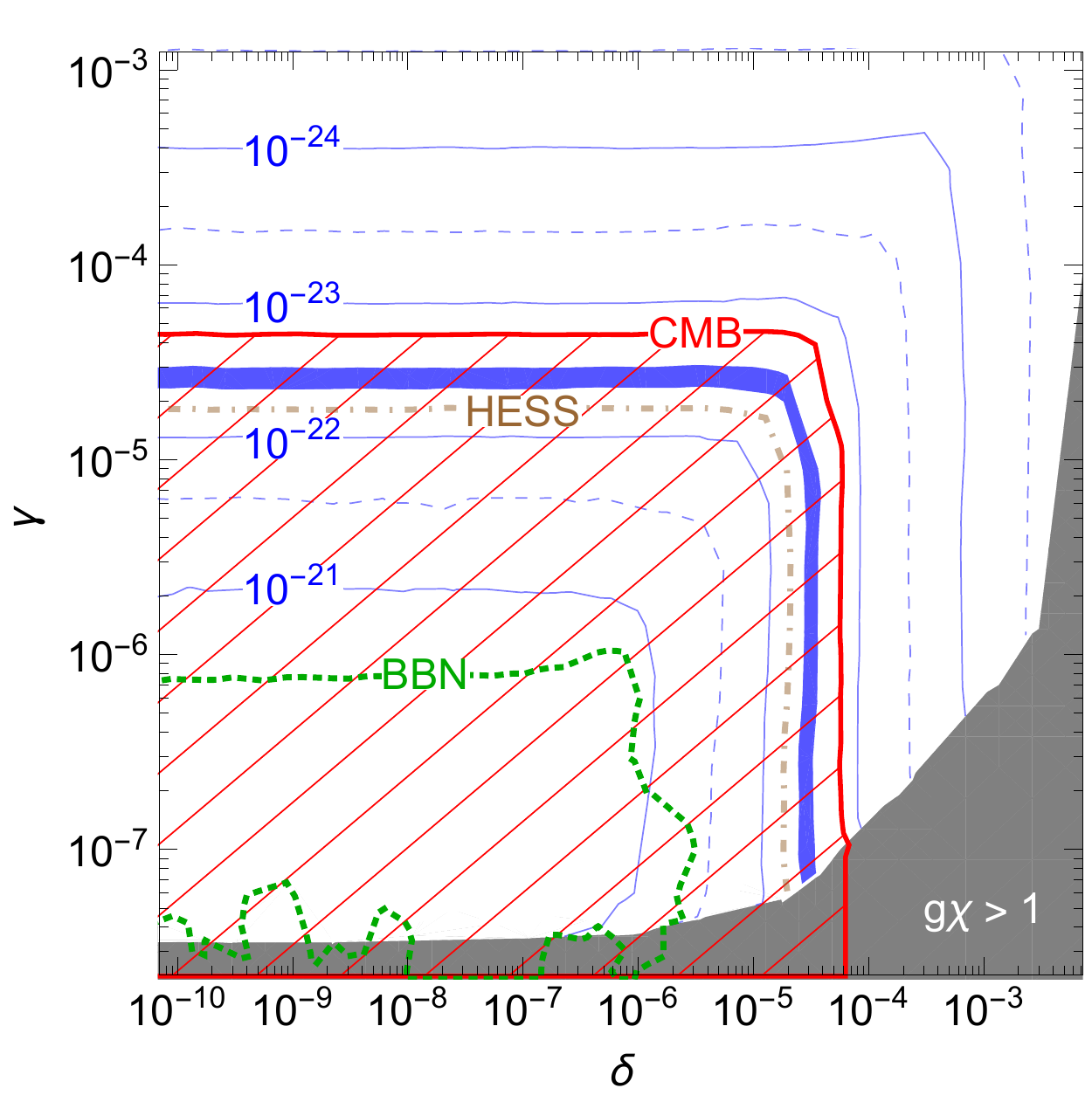}\ 
	\includegraphics[width=0.45\textwidth]{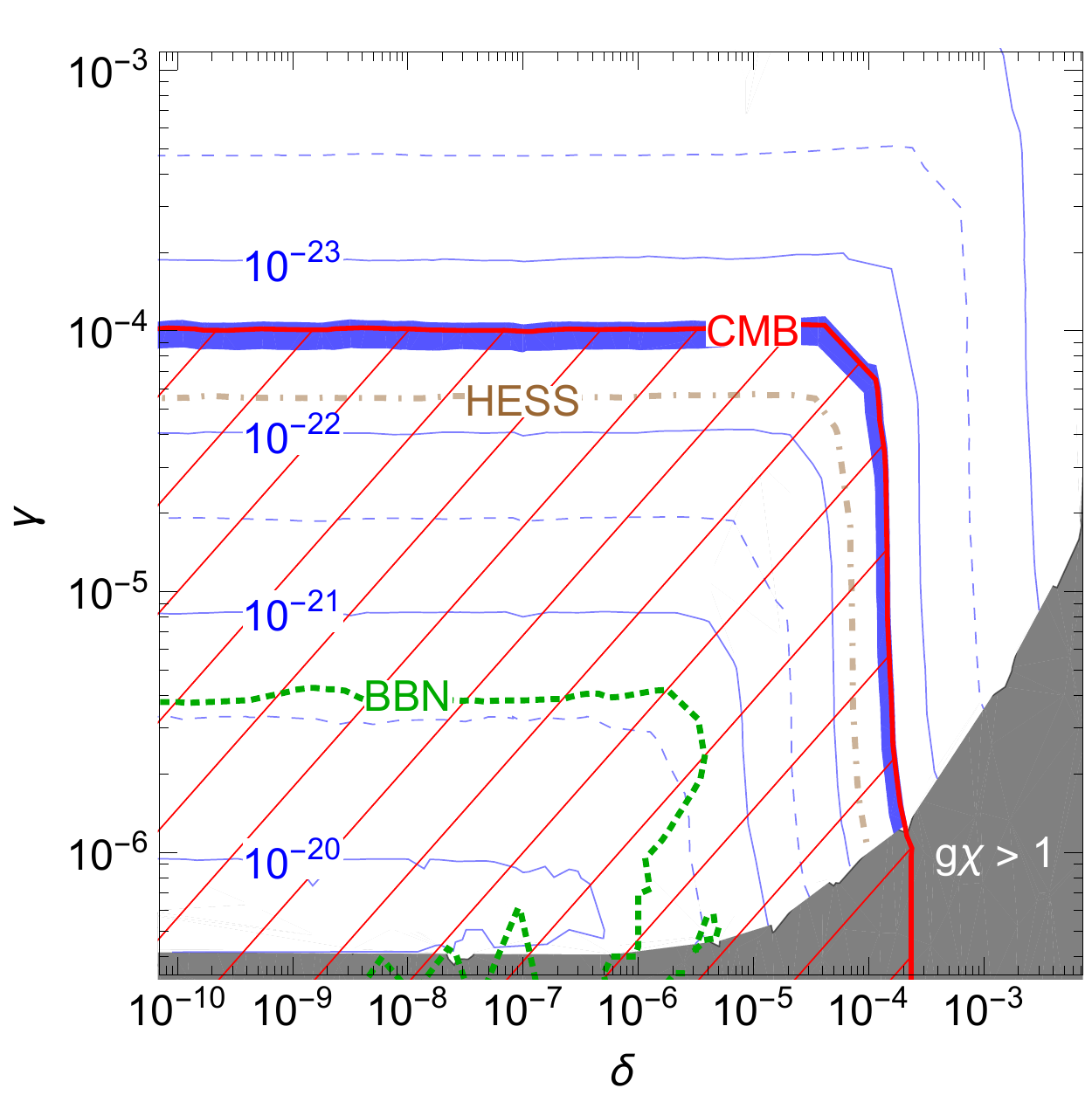}
	\caption{SA cross-sections in the galaxy today after imposing the relic density constraint, in the $(\delta\,$--$\,\gamma)$ plane, for the sequential two-body (left) and three-body (right) decay modes.  The blue contours are labelled in cm$^3$\,s$^{-1}$; the shaded blue regions explain the results of \secref{sec:fit}.  The red hatched area is excluded by CMB measurements; we also show sub-leading constraints from H.E.S.S. observations of the galactic centre (brown dot-dashed line) and BBN (green dotted line).  The grey shaded area is where the relic density can not be obtained with SA coupling $g_\chi < 1$.  The CMB constraints fully (partially) exclude the S2B (3B) preferred region, but these bounds can be weakened in the presence of a small $\mathcal{O}(1)$ substructure enhancement.  See the text for more details.}\label{fig:results}
\end{figure}

The results are plotted in \figref{fig:results} for $\sin\theta = 10^{-4}$, or $\mu_\Sigma \sim 2$\,GeV.  We found little qualitative difference for other values of $\sin\theta < 10^{-3}$, beyond two restrictions on the parameter space.  There is a lower bound on $\gamma$ coming from the decay $\Phi \to \Sigma H^\dagger$,
\begin{equation}
	\gamma \gtrsim \frac{\mu_\Sigma^2}{8\pi m_\Phi^2} \sim \frac{m_\Phi^2 - m_\Sigma^2}{4\pi v_h^2} \, \sin^2\theta \sim 5 \, \sin^2 \theta\,.
\end{equation}
This explains the region plotted in \figref{fig:results}, and reduces the parameter space for smaller mixing angles.  For larger mixing angles, the parameter space does not increase as values of $\gamma \lesssim 10^{-8}$ require $g_\chi > 1$ to reproduce the correct relic density.  Our choice of mixing angle then is roughly maximal in the $\{\delta, \gamma\}$ plane.  Aside from these effects, $\sin\theta$ only changes the rate for $\chi\chi\to\Sigma^\dagger H$, which for $\sin\theta \lesssim 10^{-3}$ is always subdominant; and so the signal cross section is independent of $\theta$ and the relic density only weakly dependent on it.

The most important conclusion from \figref{fig:results} is that the annihilation today can reach values as large as $10^{-20}\,$cm$^3\,$s$^{-1}$ while still reproducing the observed relic density.  This represents an enhancement of five orders of magnitude over the thermal relic cross section for SADM, significantly larger than was found in \refcite{1106.6027} for annihilating DM.  The cross sections of \secref{sec:fit} are easily obtainable, and are shown in the blue shaded bands.  As expected from \modeqref{eq:actualBWE}, the SA cross section is approximately set by $\max \, (\delta, \gamma)$ and hits its maximal value for $\delta \lesssim 10^{-6}$.  The contribution of off-shell $\Phi \to \chi\chi$ to the imaginary part of the two-point function leads to the optimal width $\gamma \sim 10^{-7}$ and $g_\chi \lesssim 0.1$.  Our condition $g_\chi < 1$ is not significantly affecting the preferred parameter space, with $g_\chi \sim 0.02$ (0.04) in most of the S2B (3B) best fit regions.

\begin{figure}
	\centering
	\includegraphics[width=0.45\textwidth]{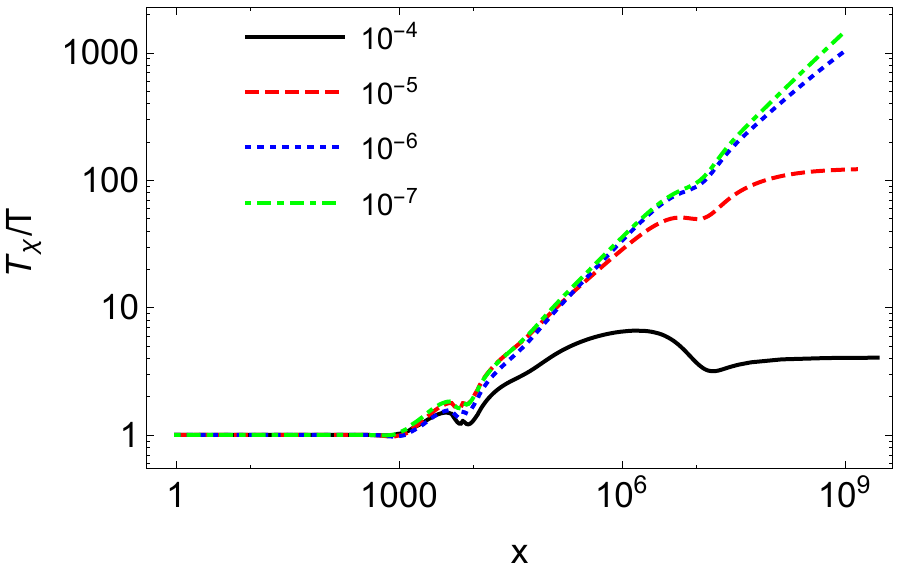}\ 
	\includegraphics[width=0.45\textwidth]{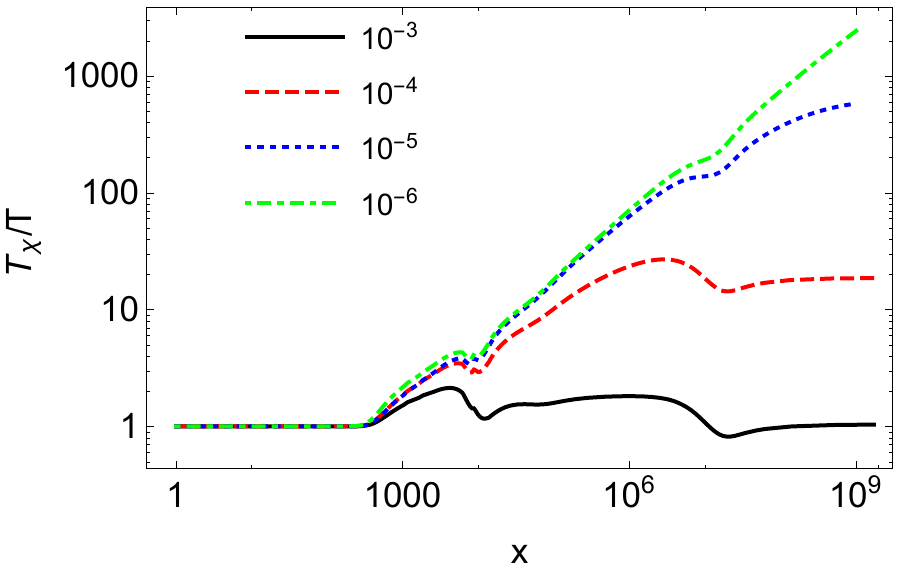}
	\includegraphics[width=0.45\textwidth]{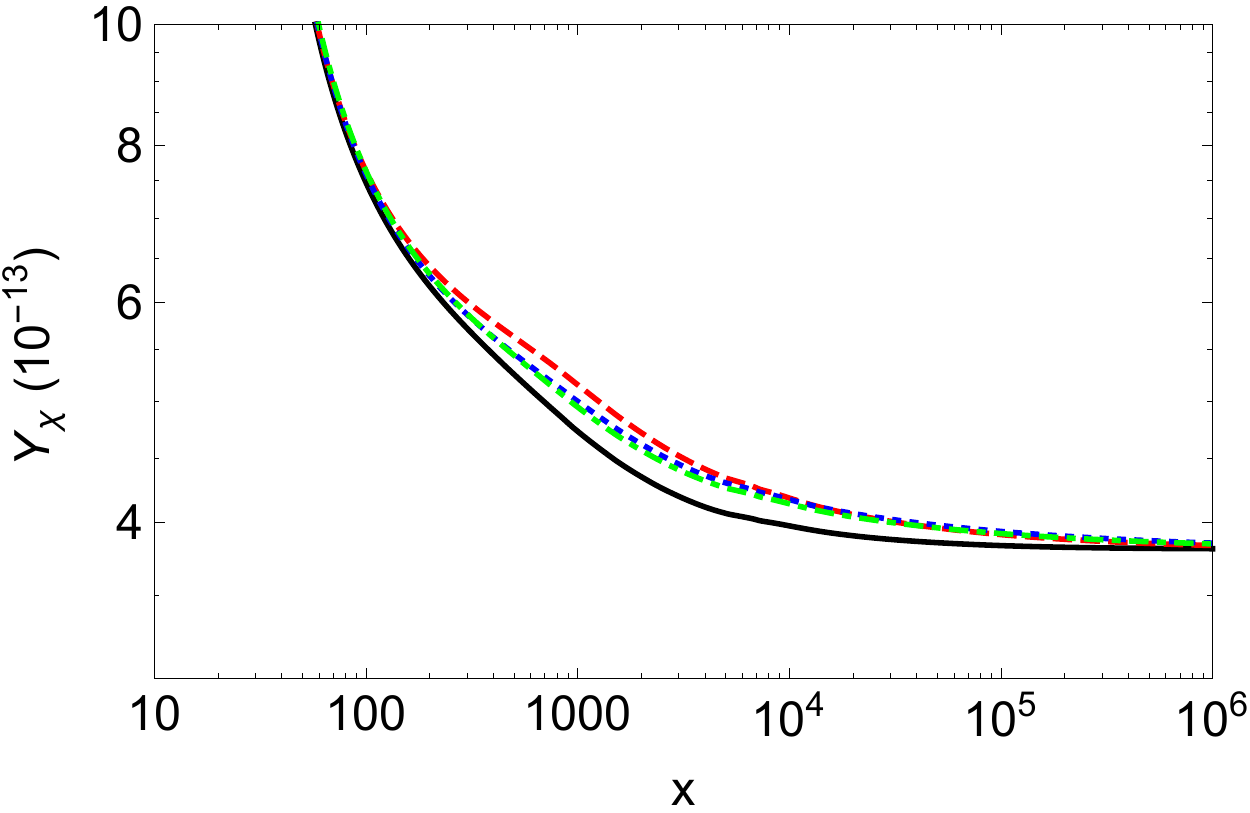}\ 
	\includegraphics[width=0.45\textwidth]{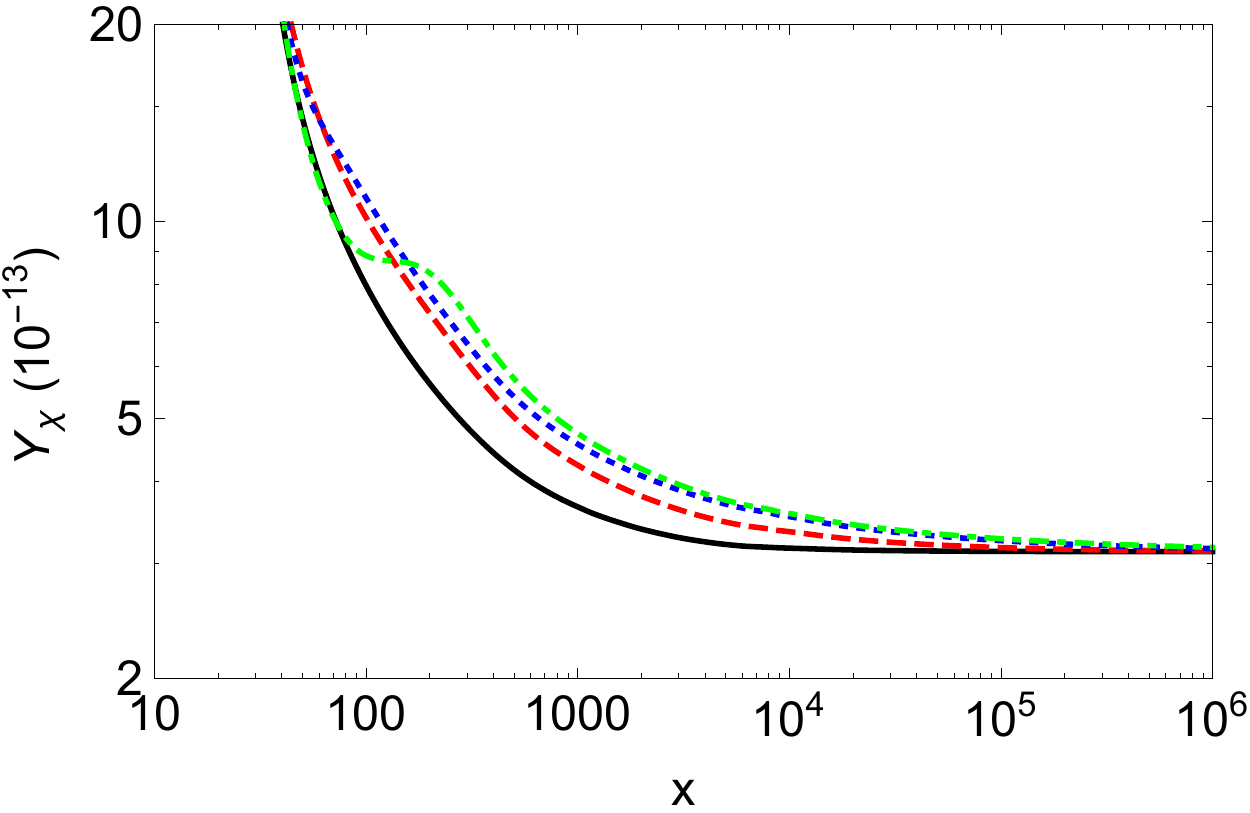}\\
	\includegraphics[width=0.45\textwidth]{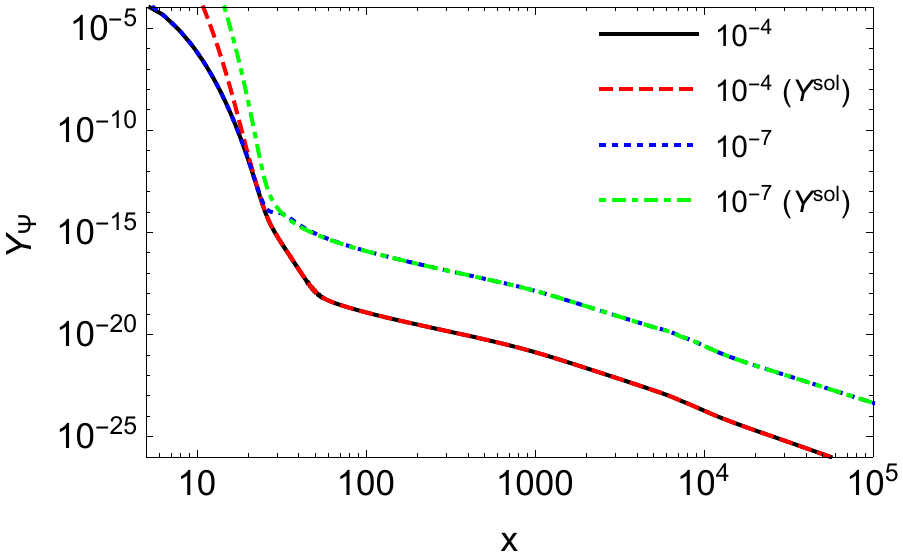}\ 
	\includegraphics[width=0.45\textwidth]{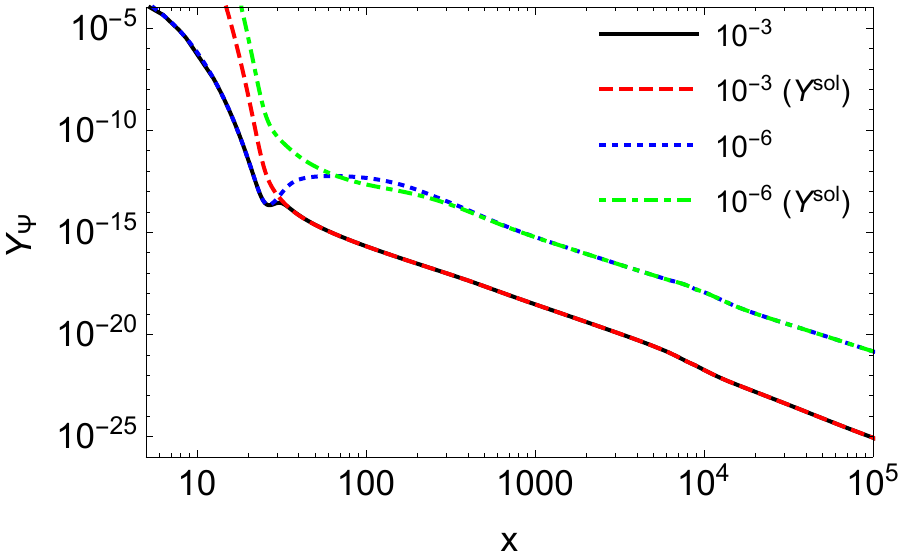}
	\caption{DM temperature $T_\chi$ (top), DM number density $Y_\chi$ (middle), and dark partner number density $Y_\Psi$ (bottom) as a function of the SM inverse temperature $x$ for the S2B (left) and 3B (right) decay modes and $\delta = 10^{-7}$.  The top two rows plot four different values of the resonance width parameter $\gamma$ as labelled in the top line.  The bottom row shows both the results of our numerical integration and the approximation of \modeqref{eq:YPsiAppr2} for two values of $\gamma$.}\label{fig:YDM}
\end{figure}

To help understand the effects of SA and kinetic decoupling, we show the evolution of $T_\chi$, $Y_\chi$, and $Y_\Psi$ in \figref{fig:YDM}.  In these plots we fix $\delta = 10^{-7}$ and vary $\gamma$ as labelled; all other parameters are as described above.  The DM temperature relative to the SM is shown in the top row of that figure, and its behaviour is simple: $T_\chi = T$ till scattering ceases to be efficient around $x \lesssim 10^3$, after which the self-heating effect of SA increases the temperature ratio ($T_\chi$ continues to decrease, just more slowly than $T$).  Smaller values of $\delta$ correspond to larger SA cross sections and so more warming, but also a higher asymptotic temperature ratio: models with smaller SA cross sections also reach their final value sooner.  The obvious features at $x \sim 10^4$ and $10^7$ in each line are SM physics: the QCD phase transition and $e^+e^-$ annihilation at $T \approx 511$\,keV, both of which warm the SM and so lower $T_\chi/T$.

The effect this has on $Y_\chi$, shown in the middle row of \figref{fig:YDM}, is relatively simple.  Since $T_\chi$ does not undergo the rapid decrease discussed in \secref{sec:bw} for conventional DM after kinetic decoupling, SA remains effective till $x \sim \min (\gamma^{-1}, \delta^{-1})$.  The relatively high DM temperature means semi-annihilation is efficient for a long time, but the enhancement is not \emph{too} great and we obtain the correct relic density for larger $y_\chi$ than would be true without self-heating.  The only notable feature occurs for $\gamma = 10^{-6}$ and the 3B decay mode, where $Y_\chi$ is approximately constant for $100 \lesssim x \lesssim 300$.  This feature can be understood by examining the evolution of $Y_\Psi$, shown in the bottom row of \figref{fig:YDM}.  We see that in all cases, the dark partner number density rapidly approaches the approximate value of \modeqref{eq:YPsiAppr2} after $Y_\chi$ departs from its equilibrium value.  Small values of $\gamma$ correspond to small dark partner widths, increasing this asymptotic value, while the width is also larger in the 3B decay mode.  In some cases, $\Psi$ must actually be replenished above its equilibrium value by SA.  For the specific case of $\gamma = 10^{-6}$ and the 3B decay mode, this increases $Y_\Psi$ to a sufficiently large value that the reverse process, $\bar{\Psi}\mu \to \chi\chi$, becomes important again and slows the depletion of DM.  This effect reduces the time available for SA to deplete the DM number density, requiring larger values of $y_\chi$ to compensate, and explains the pseudo-lower bound on $\gamma$ discussed above.

The fact that $Y_\Psi$ is well-approximated by \modeqref{eq:YPsiAppr2} for nearly all its evolution supports our use of \modeqref{eq:Ysig} for $Y_\Sigma$, rather than including an additional Boltzmann equation.  As discussed in \secref{sec:DP}, we expect our approximations to be better for states which decay more rapidly.  One possible concern is that since SA remains valid till $x \sim 10^6$, we can no longer assume the dark partner will be in kinetic equilibrium with the SM, despite the very narrow widths we consider.  Since $\Psi$-SM scattering transfers energy from the dark sector, it has the effect of \emph{lowering} $T_\chi$ and thus \emph{decreasing} the SA cross section.  Therefore if $\Psi$ also kinetically decouples from the SM before $\chi$ freeze out, we would need even larger values of $y_\chi$ to match the relic density.  Our approach is in this sense a conservative estimate of the possible enhancement.

We also consider a number of possible constraints, of which the most robust are from the cosmic microwave background (CMB).  Annihilation to electromagnetically charged states before or during recombination increases the ionisation fraction of the universe, suppressing the power spectrum at small angular scales and enhancing the polarisation power spectrum~\cite{1201.3939,astro-ph/0503486}.  The observations of these quantities then place bounds on SA at that epoch.  The cross section then need not equal its present value; specifically, DM was colder then resulting in smaller $\tilde{\gamma}$ and larger cross sections.  We include this effect, though the small values of $g_\chi$ in our best-fit regions make this a minor effect.  We use the results of \refcite{1512.08015} based on the Planck~2015 data~\cite{1507.02704}, which exclude complex DM annihilating to muons at 95\% if $\sigma v/m_\chi > 5.8 \times 10^{-27}\,$cm$^3$\,s$^{-1}$\,GeV$^{-1}$.  The limit is weaker for our model as only 39\% (35\%) of the collision energy is transmitted to the muons for the S2B (3B) case.  This still excludes all the preferred S2B parameter region; this can be reopened with a small additional substructure enhancement, at the cost of model elegance.  The 3B case is not completely excluded, due to the smaller preferred cross section and that less of the collision energy being mediated to muons.  There are also CMB bounds from the other SA channel, $\chi\chi \to \Sigma H$, as well as direct annihilation.  The former are suppressed by the smaller cross section and that even less energy is deposited into the visible sector, and for $\sin\theta = 10^{-4}$ are an order of magnitude smaller.  The annihilation bounds are negligible as the cross section is orders of magnitude weaker than for SA.

An additional early-Universe constraint comes from big-bang nucleosynthesis (BBN).  DM (semi)-annihilations can modify the predictions for the primordial elemental abundances; in particular, leptonic final states can induce photo-dissociation of $^4$He~\cite{astro-ph/0408426}, leading to stronger bounds than the CMB constraints in some models.  \Refcite{1102.4658} studied the effect of DM with Breit-Wigner-enhanced cross sections, and found an approximate bound of $\sigma v \lesssim 5\times 10^{-22}$\,cm$^3$\,s$^{-1}$ for real DM annihilating at $T_{SM} \approx 10^{-4}$\,MeV.  Including a factor of 2 for complex DM, the fraction of the collision energy transmitted to the muons, and the post-decoupling evolution of $T_\chi$, we get the constraints shown by the green dotted line in \figref{fig:YDM}. 

The next most important constraints are those from galactic centre annihilation.  In particular, cuspy DM profiles lead to bounds on TeV-scale DM from the H.E.S.S. telescope that are only slightly weaker than the thermal relic cross section~\cite{1607.08142}.  These exclude all the parameter space plotted in \figref{fig:results}, hence our focus on cored density profiles in \secref{sec:fit}.  The bounds on complex DM annihilating to light quarks for a cored NFW profile are $6 \times 10^{-23}$cm$^3$\,s$^{-1}$\,GeV$^{-1}$~\cite{1502.03244}, which we use as a conservative estimate for the bounds on annihilation to muons.  The resultant limits are weaker than those from the CMB, and also do not reach the best-fit regions.  Future bounds from CTA are likely to improve constraints by up to an order of magnitude~\cite{1408.4131}.

Finally, we briefly comment on direct and collider searches.  These are to a good approximation independent of $\delta$ and $\gamma$.  DM-nucleon scattering arises at one loop through Higgs, $Z$, and photon penguins.  We follow \refcite{1803.05660} and find that the photon-mediated interaction dominates, leading to a differential cross section~\cite{1503.03382}
\begin{equation}
	\frac{d\sigma}{dE_R} \approx \frac{\alpha^2}{4\pi} \, \mu_\chi^2 \, Z^2 \biggl( \frac{1}{E_R} - \frac{m_A}{2\mu_{\chi A}^2 v^2} \biggr) \, F_{SI}^2 (E_R) + \frac{m_A}{2\pi v^2} \, A_{eff}^2 \, F_{SI}^2 (E_R) \,.
\label{eq:DDscat}\end{equation}
Here, $E_R$ is the recoil energy, $v$ the DM incident velocity, $m_A$ the nucleon mass, $\mu_{\chi A}$ the DM-nucleon reduced mass, $F_{SI}$ the nuclear form factor, and we have neglected the spin-dependent term.  The DM magnetic moment $\mu_\chi$ and charge radius coupling $A_{eff}$ are approximately
\begin{align}
	\mu_\chi & \approx \frac{g_\Sigma^2}{8 m_\chi} \, \biggl( 1 - \frac{m_\Sigma^2}{m_\chi^2} \, \log \biggl[ 1 - \frac{m_\chi^2}{m_\Sigma^2} \biggr] \biggr) \sim 10^{-4} \, \text{GeV}^{-1} \,, \\
	A_{eff} & \approx \frac{\alpha \, g_\Sigma^2 \, Z}{12\pi \, (m_\Sigma^2 - m_\chi^2)} \, \log \biggl[ \frac{m_\Sigma}{m_\mu} \biggr] \sim Z \, 10^{-9} \, \text{GeV}^{-2} \,.
\end{align}
The two contributions to \modeqref{eq:DDscat} are similar in magnitude, though the second term is always larger.  We can make a simple estimate of the constraints by integrating the differential cross section using the Helm form factor~\cite{Helm:1956zz} and averaging over a Maxwell-Boltzmann DM velocity distribution~\cite{1708.04630}.  This gives a per-nucleon cross section for Xenon of $\Sigma_N \approx 1.6 \, (1.8) \times 10^{-49}$\,cm$^2$ in the S2B (3B) case, three orders of magnitude below the current Xenon~1T limit of $2\times10^{-46}$\,cm$^2$~\cite{1705.06655}.  The analysis of \refcite{1503.03382} that includes the full effect of the photon-mediated recoil spectrum suggests that there might be sensitivity for the full data set of Xenon~1T; otherwise we must wait for next-generation experiments such as DARWIN~\cite{1606.07001}.

Since all our dark sector particles are at the TeV-scale, only the LHC has reach for direct production.  The absence of any new coloured particles results in no sensitivity.  The scalar $\Sigma^+$ behaves very similarly to a smuon, being produced through electroweak processes and decaying to a muon plus missing energy.  The current bounds from CMS and ATLAS are $m_\Sigma \gtrsim 450$\,GeV~\cite{CMS:2017mkt} and 350\,GeV~\cite{1403.5294}, respectively.  The bounds on the dark partner are weaker, due to its visible decay products being softer.  Precision observables in the lepton sector can also be highly constraining; in particular, flavour-changing neutral currents essentially force us to consider a $\chi$-$\Sigma$ Yukawa coupling only to muons, with zero tau or electron component.  The next most interesting observable is the muon anomalous magnetic moment, where the one-loop contribution is~\cite{1803.05660,0801.1826}
\begin{equation}
	a_\mu = - \frac{g_\Sigma^2}{32\pi^2} \, \frac{m_\mu^2}{m_\Sigma^2} \, f \biggl( \frac{m_\chi^2}{m_\Sigma^2} \biggr) \approx 
	\begin{cases}
		-1.75 \times 10^{-12} & \text{S2B,} \\
		-9.57 \times 10^{-13} & \text{3B,}
	\end{cases}
\end{equation}
where
\begin{equation}
	f(x) = \frac{1 - 6x + 3x^2 + 2x^3 - 6x^2 \log (x)}{6(1-x)^4} \,.
\end{equation}
This has the wrong sign to explain the discrepancy~\cite{Patrignani:2016xqp} $\Delta a_\mu = a_\mu^{exp} - a_\mu^{SM} = 288 \pm 63 \pm 49 \times 10^{-11}$.  However, the additional term is smaller than the uncertainty on $\Delta a_\mu$ and so our model is no less compatible than the SM.

%% file: Files/conc.tex
\section{Conclusions}\label{sec:conc}

Semi-annihilation is a generic feature of particle dark matter phenomenology that is resistant to searches at colliders and direct detection experiments.  With cosmic ray searches being the primary tool to explore SADM phenomenology today, it is important to study the range of potential signals that can exist.  In this work, we have considered models where SA is enhanced at low temperatures, allowing for greater cross sections today than during thermal freeze-out.  Such models are of obvious interest, as they are likely to be the most easiest to test and thus the first to be discovered or excluded.

This paper may be divided into two parts.  The first consists of \secsref{sec:DP}, \ref{sec:TempEvo} and~\ref{sec:bw}, where we discussed general features of SADM phenomenology.  We began in \secref{sec:DP} by reviewing the concept of \emph{dark partners}, first introduced in \refcite{1611.09360}.  These are states charged under both dark and visible sector symmetries, that are not DM but can appears as final states in DM-initiated SA processes.  This allows $2\to 2$ SA with arbitrary SM final states.  In addition to presenting the relevant Boltzmann equations for dark partners for the first time, we derived useful analytic approximations and asymptotic solutions given in \modeqsref{eq:YPsiAppr2} and~\eqref{eq:YPsiInf}.

However, our most important general results are found in \secref{sec:TempEvo} (with additional details in \appref{app:fullBE}), where we discuss the thermal evolution of SADM.  We generalise earlier results from \refcite{1707.09238} to include models with dark partners, and discuss the asymptotic evolution of the DM temperature, $T_\chi$.  SA converts mass into DM kinetic energy, \emph{self-heating} the dark sector.  Consequently, after DM-SM scattering ceases to be efficient, $T_\chi$ redshifts as \emph{radiation}, not matter, despite remaining highly non-relativistic.  We clarified the generality of this result, particularly how it applies to dark sectors with dark partners.  Most interestingly, when the SA cross section is enhanced at late times, it is possible to have the dark sector \emph{hotter} than the SM, at least during the radiation-dominated epoch, as shown in \modeqref{eq:dydxSAasymp}.

In \secref{sec:bw} we consider the implications of this temperature evolution on one specific type of enhanced cross section, where there is an $s$-channel resonance  nearly on-threshold.  Annihilating DM has a \emph{theoretical} bound on the possible enhancement of $\mathcal{O}(10^2)$ that arises from early kinetic decoupling.  This does not apply to SADM due to this self-heating effect, potentially expanding the available model space.

The second part of this paper applied these general results to a case study, a specific model using a Breit-Wigner resonance to explain the positron excess seen by AMS-02 and earlier experiments.  We first describe this model in \secref{sec:model}, with a possible UV motivation given in \appref{app:UV}.  This model has two important phases, according to the different decay modes of the dark partner, either a direct three-body or sequential two-body decay.  In \secref{sec:fit} we fit the model to the AMS-02 positron data.  As expected, the required cross sections are orders of magnitude larger than the thermal relic cross section.  The role of the different decay modes is also clearly seen in \figref{fig:results}.  Finally, we solve the Boltzmann equations and compare the best-fit regions to other experimental constraints in \secref{sec:rd}.  We see that the self-heating effect allows SA to continue to deplete the DM number density till late times, $x \sim 10^6$.  This in turn allows the correct relic density to be obtained for small widths and large enhancements; SA cross sections today at least five orders of magnitude larger than the thermal relic cross section are achievable.  We see that there are stringent constraints from CMB observations and, for cuspy DM profiles, $\gamma$-rays from the galactic centre; though these are generic constraints on any putative DM explanation of the positron excess.  We also see that, as expected, direct and collider bounds on the SADM model are currently quite weak.

In this work, we restricted our attention to a fairly narrow possibility among the model space of SADM with interactions with a non-trivial temperature dependence.  One obvious future extension is to consider other processes than an $s$-channel resonance.  Non-perturbative interactions such as the Sommerfeld effect or bound state formation are an alternative possibility where the cross section can be enhanced at low temperatures.  SA and the subsequent modification to the DM temperature could be interesting for its effect on the relic density and signals.  Another line of inquiry would be to consider a more general interplay of SA, annihilation, and scattering.  Our specific model included DM-SM scattering to maintain kinetic equilibrium, but the case where $T_\chi$ is determined by SA alone is potentially interesting.  This could also include theories where \emph{annihilation} is enhanced, and dominates the determination of the relic density, but SA is important as it governs the late-time temperature evolution.  These ideas demonstrate that the theory space for SADM remains rich and open to investigation.

%% file: Files/Derivation.tex
\section{Boltzmann equations with kinetic decoupling}\label{app:fullBE}

The general Boltzmann equation for the evolution of the phase space density of a particle species is 
\begin{equation}
	\hat{L} [f_i] = \hat{C} [f_i] \,,
\label{eq:BEgen}\end{equation}
where $\hat{L}$ is the Liouville operator encapsulating the effects of the metric (or in a cosmological sense, the expansion) and $\hat{C}$ the collision operator describing interactions.  In a FRW cosmology and taking $f_i \equiv f(E, t)$,
\begin{equation}
	\hat{L} [f_i] = E \frac{\partial f_i}{\partial t} - H \, \vec{p}\,{}^2 \, \frac{\partial f_i}{\partial E} \,,
\label{eq:FRWLO}\end{equation}
with $H$ the Hubble rate.  Rather than deal with \modeqref{eq:BEgen} directly, we deal with its moments, converting an evolution equation for a function of momentum into one for scalar functions of time (only).  In particular, we note that 
\begin{align}
	n_i & = g_i \int \frac{d^3 p}{(2\pi)^3} \, f_i(E)\,,  & T_i = \frac{P_i}{n_i} & = \frac{g}{3} \int \frac{d^3 p}{(2\pi)^3} \, \frac{\vec{p}\,{}^2}{E} \, f_i(E) \,,
\end{align}
where $g$ is the number of internal degrees of freedom, $P$ is the pressure, and in the second relation we used the ideal gas law.  Let us define the dimensionless parameters
\begin{equation}
	Y_i = \frac{n_i}{s} \,, \qquad y_i = \frac{m_i \, T_i}{s^{2/3}} \,, \qquad x = \frac{m_0}{T} \,, \qquad Z = \biggl(1 - \frac{x g_{\ast S}' (x)}{3 g_{\ast S} (x)} \biggr)^{-1} \,,
\end{equation}
where $s$ is the entropy density, $T$ the SM temperature, $m_0$ any reference mass, and $g_{\ast S} (x)$ the effective number of relativistic degrees of freedom.  Then we derive the standard results
\begin{gather}
	x s H Z \, \frac{d Y_i}{dx} = g_i \int \frac{d^3 p}{(2\pi)^3} \, \frac{1}{E} \, \hat{L} [f_i] = g_i \int \frac{d^3 p}{(2\pi)^3} \, \frac{1}{E} \, \hat{C} [f_i] \,, \label{eq:defdYdx}\\
	\frac{3 s^{5/3} x H Z}{m_i} \biggl( y_i \, \frac{dY_i}{dx} + Y_i \, \frac{dy_i}{dx} \biggr) - s H Y_i \biggl\langle \frac{p^4}{E^3} \biggr\rangle = g_i \int \frac{d^3 p}{(2\pi)^3} \, \frac{\vec{p}\,{}^2}{E^2} \, \hat{L} [f_i] = g_i \int \frac{d^3 p}{(2\pi)^3} \, \frac{\vec{p}\,{}^2}{E^2} \, \hat{C} [f_i] \,, \label{eq:defdytdx}
\end{gather}
where the first equality in each line follows from the FRW Liouville operator, \modeqref{eq:FRWLO}, and the second trivially by \modeqref{eq:BEgen}.  Comparing \modeqref{eq:defdytdx} to \modeqref{eq:dytgen}, we see that the operators $C_2$ introduced in the latter are given by
\begin{equation}
	C_2 = \frac{g_i}{3 m_i T_i n_i} \int \frac{d^3 p}{(2\pi)^3} \, \frac{\vec{p}\,{}^2}{E^2} \, \hat{C} [f_i] \,.
\end{equation}

The collision operator includes terms for all possible processes.  However, only number changing processes have non-zero contribution to \modeqref{eq:defdYdx} (so DM-SM and DM-DM scattering vanish), and only processes that change the total DM kinetic energy contribute to \modeqref{eq:defdytdx} (so DM-DM scattering is zero).  The general form for the integral in \modeqref{eq:defdYdx} is~\cite{Kolb:1990vq}
\begin{multline}
	g_i \int \frac{d^3 p}{(2\pi)^3} \, \frac{1}{E} \, \hat{C} [f_i] = - \int \biggl( \prod_{\alpha = i,j,k,\ldots} d\Pi_\alpha \biggr) \, (2\pi)^4 \, \delta^{(4)} \Bigl(\sum p\Bigr) \sum \lvert \mathcal{M} (ij \ldots \to k\ldots) \rvert^2 \\
	\times \bigl( f_i(p_i) \, f_j(p_j) \ldots (1 \pm f_k (p_k) ) \ldots - f_k (p_k) \ldots (1 \pm f_i (p_i) ) (1 \pm f_j (p_j)) \ldots \bigr) \,.
\label{eq:intCO}\end{multline}
We have introduced the Lorentz-invariant momentum integral including a factor of the internal degrees of freedom, $d\Pi_\alpha = g_\alpha d^3 p_\alpha/((2\pi)^3 2E_\alpha)$.  The integral in \modeqref{eq:defdytdx} contains an additional factor of $\vec{p}\,{}^2/E$, as is clear from the definition.  Making the assumptions that $CP$ is conserved, the phase space functions can be approximated as Maxwellian, and that $f_\alpha \ll 1$ for all species leads to the usual results for the Boltzmann equation for $Y_i$ in terms of thermally averaged (co)-annihilation cross sections.  Specifically, with these assumptions the first combination of phase space factors in the second line of \modeqref{eq:intCO} depends only on the the initial (dark sector) momenta.  The second depends only on the total centre-of-mass frame energy, and can be rewritten using conservation of energy in terms of the equilibrium phase space factors for the initial state particles.  All these terms factor out, so that
\begin{align}
	g_i \int \frac{d^3 p}{(2\pi)^3} \, \frac{1}{E} \, \hat{C} [f_i] & = - \int d\Pi_i \, d\Pi_j \, \ldots \bigl( f_i(p_i) \, f_j(p_j) \ldots - f_i^{eq} (p_i) \, f_j^{eq} (p_j) \ldots \bigr) \notag \\
	& \quad \times \int d\Pi_k \, d\Pi_l \, \ldots \, (2\pi)^4 \, \delta^{(4)} \Bigl(\sum p\Bigr) \, \sum \lvert \mathcal{M} (ij \ldots \to k\ldots) \rvert^2  \,. \label{eq:factorint}
\end{align}
The second line is proportional to the cross section or decay rate multiplied by some kinetic factors.  The overall expression then takes the usual form; for example, for $2 \to n$ processes defining 
\begin{equation}
	 \langle \sigma v (\chi\chi^\dagger \to SM) \rangle_{neq} = \frac{g_\chi^2}{n_\chi^2} \int \frac{d^3 p_\chi}{(2\pi)^3} \, \frac{d^3 p_{\chi^\dagger}}{(2\pi)^3} \, f_\chi (p_\chi) \, f_\chi (p_{\chi^\dagger}) \, \sigma (\chi\chi^\dagger \to SM) \, \frac{\sqrt{(p_\chi \cdot p_{\chi^\dagger})^2 - m_\chi^4}}{E_\chi \, E_{\chi^\dagger}} \,,
\label{eq:svneq}\end{equation}
and $\langle \sigma v \rangle_{eq}$ using the equilibrium functions $f_\chi^{eq}$,  obviously implies \modeqref{eq:dYwithTchi} even in the presence of kinetic decoupling, $T_\chi \neq T$.  In practice we usually  compute \modeqref{eq:svneq} assuming that $f_\chi$ has a Maxwellian form, or physically that $\chi$ self-scattering remains efficient till after kinetic decoupling.  The contribution to \modeqref{eq:defdytdx} from annihilation can be done in the same way, giving \modeqsref{eq:CollTAnn} and~\eqref{eq:AnnytTA}.  The scattering term is more complicated, since these assumptions do not apply; see \refcite{phdscatter}, but the final result is given by \modeqsref{eq:C2scatter} and~\eqref{eq:gammascatter}.

For semi-annihilation, the only relevant difference is that both the initial and final states contain a dark sector particle.  However, when $T_\chi \neq T$ this prevents us from factoring the integral as in \modeqref{eq:factorint}.  Let us first consider the case where the only external dark sector particles are dark matter.  The contribution to \modeqref{eq:defdYdx} instead takes the form
\begin{align}
	& g_i \int \frac{d^3 p}{(2\pi)^3} \, \frac{1}{E} \, \hat{C} [f_i] = \notag \\
	& - \int d\Pi_i \, d\Pi_j \, \ldots \, f_i(p_i) \, f_j(p_j) \ldots \int d\Pi_k \, d\Pi_l \, \ldots \, (2\pi)^4 \, \delta^{(4)} \Bigl(\sum p\Bigr) \, \sum \lvert \mathcal{M} (ij \ldots \to kl\ldots) \rvert^2 \notag \\
	& + \int d\Pi_k \, d\Pi_l \, \ldots \, f_k(p_k) \, f_l(p_l) \ldots \int d\Pi_i \, d\Pi_j \, \ldots \, (2\pi)^4 \, \delta^{(4)} \Bigl(\sum p\Bigr) \, \sum \lvert \mathcal{M} (kl \ldots \to ij\ldots) \rvert^2 \,.
\end{align}
The second (third) line is a thermally averaged cross sections of the forward (reverse) process.  In particular, if we consider the $2\to 2$ case $\chi\chi \to \chi^\dagger \phi$, with $\phi$ a visible-sector particle, then this reproduces \modeqsref{eq:dYSA1Tchi} and~\eqref{eq:InvSVavg}.

When computing the effect of this type of SA on the DM temperature, we must include the contribution on the kinetic energy of both the initial and final states.  To be concrete, let us focus on precisely the process $\chi\chi \to \chi^\dagger \phi$; then
\begin{align*}
	C_2 & = - \frac{1}{6m_\chi T_\chi n_\chi} \int d\Pi_i d\Pi_j f_i f_j \int d\Pi_k d\Pi_l (2\pi)^4 \delta^{(4)} \Bigl(\sum p\Bigr) \sum \lvert \mathcal{M} (ij \to kl) \rvert^2 \biggl( \frac{\vec{p}_i{}^2}{E_i} + \frac{\vec{p}_j{}^2}{E_j} - \frac{\vec{p}_k{}^2}{E_k} \biggr) \\
	& \quad + \frac{1}{6m_\chi T_\chi n_\chi} \int d\Pi_k d\Pi_l f_k f_l \int d\Pi_i d\Pi_j (2\pi)^4 \delta^{(4)} \Bigl(\sum p\Bigr) \sum \lvert \mathcal{M} (kl \to ij) \rvert^2 \biggl( \frac{\vec{p}_i{}^2}{E_i} + \frac{\vec{p}_j{}^2}{E_j} - \frac{\vec{p}_k{}^2}{E_k} \biggr) .
\end{align*}
Though we have written the $\vec{p}_{i,j}$ terms separately, their contribution is equal by symmetry.  If we define
\begin{equation}
	\mathcal{U}_{ij \to kl} (p_i, p_j) = \frac{1}{\sigma (ij \to kl)} \, \frac{1}{4\sqrt{(p_i \cdot p_j)^2 - m_i^2 m_j^2}} \int d\Pi_2 \, \sum \lvert \mathcal{M} (ij \to kl) \rvert^2 \frac{\vec{p}_k{}^2}{m_kE_k} \,,
\label{eq:defP}\end{equation}
then we can write
\begin{align}
	C_2 & = - \frac{n_\chi}{m_\chi} \biggl( \langle \sigma v (\chi\chi \to \chi^\dagger \phi) \rangle_{2,neq} + \frac{m_\chi}{6T_\chi} \, \langle \sigma v (\chi\chi \to \chi^\dagger \phi) \mathcal{U}_{\chi\chi \to \chi^\dagger \phi}\rangle_{neq} \biggr)\notag \\
	& \quad - \frac{n_\phi^{eq}}{m_\chi} \biggl( \frac{m_\chi}{3T_\chi} \, \langle \sigma v (\chi^\dagger \phi \to \chi\chi) \mathcal{U}_{\chi^\dagger \phi \to \chi\chi} \rangle_{neq} - \langle \sigma v (\chi^\dagger \phi \to \chi\chi) \rangle_{2,neq} \biggr) \,. \label{eq:C2SA1}
\end{align}
This expression is exact and (unlike those in \refscite{1707.09238,1805.05648}) defined for any phase space distribution.  It can also be extended to $2\to n$ processes with an appropriate change to \modeqref{eq:defP}.

Before discussing some simplifications of \modeqref{eq:C2SA1}, let us comment on the two terms with factors of $m_\chi/T_\chi$ that become large at low temperatures.  For the third term in \modeqref{eq:C2SA1}, this enhancement is not realised.  The reverse process $\chi\phi^\dagger \to \chi\chi$ requires the initial states to have a minimum kinetic energy to proceed, and as such will be exponentially suppressed by the phase space factors at late times.  In contrast, the second term in \modeqref{eq:C2SA1} \emph{can} proceed at threshold and has no such suppression.  Indeed, in the non-relativistic limit $\mathcal{U}_{\chi\chi \to \chi^\dagger \phi}$ asymptotes to a constant,
\begin{equation}
	\mathcal{U}_{\chi}^S \equiv \mathcal{U}_{\chi\chi \to \chi^\dagger \phi} (0, 0) = \frac{(9 m_\chi^2 - m_\phi^2)(m_\chi^2 - m_\phi^2)}{4m_\chi^2 \, (5m_\chi^2 - m_\phi^2)} \,.
\end{equation}
Physically, this is the effect discussed in \secref{subsec:TEasymp}, namely that SA provides a roughly fixed kinetic energy to the final state particle which becomes large compared to the thermal kinetic energy at late times.

While \modeqref{eq:C2SA1} is exact, it is not the most convenient form for calculation or interpretation.  In practice, we usually assume that the dark matter phase space distribution is Maxwellian, \emph{i.e.} that DM self-scattering remains efficient, even after DM-SM scattering decouples.  In that case, we can use \modeqsref{eq:AltPhSp} and~\eqref{eq:defDelta} to convert the reverse cross sections into forward ones:
\begin{align}
	\frac{1}{n_\chi} & \int d\Pi_k d\Pi_l \, f_k f_l \int d\Pi_i d\Pi_j \, (2\pi)^4 \delta^{(4)} \Bigl(\sum p\Bigr) \, \sum \lvert \mathcal{M} (kl \to ij) \rvert^2 \, \biggl( \frac{2 \vec{p}_i{}^2}{E_i} - \frac{\vec{p}_k{}^2}{E_k} \biggr) \notag \\
	& = \frac{1}{n_\chi^{eq} (T_\chi)} \int d\Pi_i d\Pi_j \, f_i^{eq} f_j^{eq} \notag \\
	& \quad \times \int d\Pi_k d\Pi_l \, (2\pi)^4 \delta^{(4)} \Bigl(\sum p\Bigr) \, \sum \lvert \mathcal{M} (ij \to kl) \rvert^2 \, \biggl( \frac{2 \vec{p}_i{}^2}{E_i} - \frac{\vec{p}_k{}^2}{E_k} \biggr) \, e^{-E_k \Delta_\chi/m_\chi} \,.
\end{align}
If we further make the non-relativistic approximation that the $\chi^\dagger$ energy and momenta in the centre-of-mass frame and cosmological frames are equal, then we can use the functions $\mathcal{S}$ and $\mathcal{S}_{\mathcal{T}}$ defined in \modeqsref{eq:defSTT} and~\eqref{eq:defS2} to straightforwardly derive \modeqref{eq:C2SA1_simp}.

For models that include dark partners, as discussed in \secref{subsec:TESA} we restrict ourselves to models where we can assume $T_\Psi = T$ during relevant eras.  The contribution of SA to the various Boltzmann equations is then simple, as the reverse process can be written in terms of the forward process at equilibrium in the usual manner:
\begin{equation}
	f_\Psi (p_\Psi) \, f_l (p_{SM}) \ldots = \frac{n_\Psi}{n_\Psi^{eq}} \, f_i^{eq} \, f_j^{eq} \,,
\end{equation}
which implies \modeqsref{eq:dYSA} and~\eqref{eq:C2SADP}.  The dark partner decay contribution in general requires using the inverse process again,
\begin{align*}
	\frac{dY_\chi}{dx} & \supset \frac{\Gamma_\Psi}{xHZ} \, Y_\Psi \, \biggl\langle \frac{m_\Psi}{E} \biggr\rangle_\Psi - \frac{1}{xsHZ} \int d\Pi_i \, d\Pi_j \ldots \, f_i \, f_j \ldots \, \frac{\pi}{2} \, \delta (E_{cm}^2 - m_\Psi^2) \, \sum \lvert \mathcal{M} (ij\ldots \to \Psi)\rvert^2 \\
	& = \frac{\Gamma_\Psi}{xHZ} \, Y_\Psi \, \biggl\langle \frac{m_\Psi}{E} \biggr\rangle_\Psi - \frac{s}{xHZ} \, Y_\chi Y_{SM} \, \langle \sigma v (\chi + SM \to \Psi) \rangle_{T_\chi, T} \,.
\end{align*}
where the average in the first term is over the $\Psi$ phase space function and is the effect of time dilation.  The thermal average of the $n \to 1$ inverse decay cross section is, for two-body decays, directly proportional to the width.  Taking $\Psi$ to be non-relativistic implies $\langle m_\Psi/E \rangle \approx 1$, while taking $f_\chi$ to be Maxwellian allows us to define the function
\begin{equation}
	\mathcal{D} (T, T_\chi) = \frac{n_\chi^{eq} (T)}{n_\chi^{eq} (T_\chi)} \, \frac{1}{2m_\Psi \Gamma_\Psi} \int d\Pi_i \, d\Pi_j \ldots \, \sum \lvert \mathcal{M} (ij\ldots \to \Psi)\rvert^2 \, e^{-E_\chi \Delta_\chi/m_\chi} \,,
\label{def:D}\end{equation}
from which \modeqref{eq:dYDPdec} follows.

To describe the effect of $\Psi$ decay on $y_\chi$ we introduce the dimensionless parameter
\begin{equation}
	\mathcal{U}^D_\chi = \frac{1}{2 m_\Psi \Gamma_\Psi} \int d\Pi_i \, d\Pi_j \ldots \, \sum \lvert \mathcal{M} (ij\ldots \to \Psi)\rvert^2 \, \frac{p_\chi^2}{m_\chi E_\chi} \,,
\label{eq:defgvchi}\end{equation}
which is the mean value of the DM Lorentz boost times velocity, in the parent rest frame.  Additionally we define the function
\begin{equation}
	\mathcal{D}_{\mathcal{T}} (T, T_\chi) = \frac{n_\chi^{eq} (T)}{n_\chi^{eq} (T_\chi)} \, \frac{1}{\overline{\gamma v}_\chi} \, \frac{1}{2m_\Psi \Gamma_\Psi} \int d\Pi_i \, d\Pi_j \ldots \, \sum \lvert \mathcal{M} (ij\ldots \to \Psi)\rvert^2 \, \frac{p_\chi^2}{m_\chi E_\chi} \, e^{-E_\chi \Delta_\chi/m_\chi} \,.
\label{def:DT}\end{equation}
Together with our usual assumptions, these allow us to write the forward decay contribution as
\begin{equation}
	C_2^{dec} \supset \frac{1}{3n_\chi T_\chi} \int d\Pi_\Psi f_\Psi \int d\Pi_i \, d\Pi_j \ldots \sum \lvert \mathcal{M} (ij\ldots \to \Psi)\rvert^2 \, \frac{p_\chi^2}{m_\chi E_\chi} \approx \mathcal{U}^D_\chi \, \frac{\Gamma_\Psi}{3T_\chi} \, \frac{n_\Psi}{n_\chi} \,,
\end{equation}
and the reverse as 
\begin{align}
	C_2^{dec} & \supset - \frac{1}{3n_\chi T_\chi} \int d\Pi_\Psi \int d\Pi_i \, d\Pi_j \ldots f_i f_j \ldots \sum \lvert \mathcal{M} (ij\ldots \to \Psi)\rvert^2 \, \frac{p_\chi^2}{m_\chi E_\chi} \notag \\
	& = - \frac{1}{3n_\chi^{eq} (T_\chi) \, T_\chi} \int d\Pi_\Psi \, f_\Psi^{eq} \int d\Pi_i \, d\Pi_j \ldots \sum \lvert \mathcal{M} (ij\ldots \to \Psi)\rvert^2 \, \frac{p_\chi^2}{m_\chi E_\chi} \, e^{-E_\chi \Delta_\chi/m_\chi} \notag \\
	& \approx - \mathcal{U}^D_\chi \, \mathcal{D}_{\mathcal{T}} (T, T_\chi) \, \frac{\Gamma_\Psi}{3 T_\chi} \, \frac{n_\Psi^{eq}}{n_\chi^{eq}} \,,
\end{align}
from which \modeqref{eq:C2decDP} follows.

%% file: Files/UV.tex
\section{Lepton-Number Model}\label{app:UV}

In \secref{sec:model} we defined our model, based on an \emph{ad hoc} global $\set{Z}_3$ symmetry and a particular flavour structure for the Yukawa couplings.  While not required, there is a theoretical preference that global symmetries derive either from the breaking of a gauge symmetry, or as accidental symmetries.  The latter possibility is not possible in our model, as the gauge content and charges given in \tabref{tab:npf} would allow a Yukawa coupling $\Phi \, \bar{\chi}\chi$, which breaks the global $\set{Z}_3$.  Additionally, there is nothing in the model as constructed that would prevent couplings to the electron and/or tau.  Couplings to the first generation would lead to a much harder positron spectrum from DM SA, while a generic flavour structure of the couplings would lead to tightly constrained contributions to FCNC operators involving electrons.

In this section we show that both these problems can be explained using a $U(1)_{L_\mu - L_\tau}$ symmetry.  Additionally, certain unnecessary terms in the scalar potential~\eqref{eq:scapot} can be forbidden, and the cubic $\mu_\Sigma$ rendered technically natural.  This symmetry is anomaly-free within the SM, and all of our new fermions are Dirac, so it can be gauged.  A gauge symmetry of this form has been used to explain the structure of neutrino masses and mixings~\cite{1607.04046,1805.04415,1805.03942,1512.04207,Baek:2001kca} and also has been invoked in other attempts to explain the positron excess~\cite{1607.04046,0903.0122,0811.1646}.

\begin{table}
	\centering
	\begin{tabular}{|c|c|c|c|}
		\hline
		Particle & Spin & $U(1)_{L_\mu - L_\tau}$ & $SU(2)_L \times U(1)_Y$ \\
		\hline
		$\chi$ & $1/2$ & $\phantom{-}1/3$ & $1_0$ \\
		$\Psi$ & $1/2$ & $\phantom{-}1/3$ & $1_{-1}$ \\
		$\Phi$ & 0 & $-2/3$ & $1_0$ \\
		$\Sigma$ & 0 & $-2/3$ & $2_{1/2}$ \\
		\hline
		\hline
		$H$ & 0 & $\phantom{-}0$ & $2_{1/2}$ \\
		$\mu_R$ & $1/2$ & $\phantom{-}1$ & $1_{-1}$ \\
		$\tau_R$ & $1/2$ & $-1$ & $1_{-1}$ \\
		$L_\mu$ & $1/2$ & $\phantom{-}1$ & $2_{-1/2}$ \\
		$L_\tau$ & $1/2$ & $-1$ & $2_{-1/2}$ \\
		\hline
		\hline
		$\mathcal{H}$ & 0 & $\phantom{-}1$ & $1_0$ \\
		\hline
	\end{tabular}
	\caption{Particle content of the lepton number dark matter model.  The top section contains the same matter as the theory of \secref{sec:model}; the middle section contains SM particles; and the last is (a possible choice for) the additional Higgs that breaks the new gauge symmetry.}\label{tab:LmLt}
\end{table}

The charges under the new gauge symmetry are given in \tabref{tab:LmLt}.  In addition to the four dark sector fields considered previously, we include the charges of the relevant SM fields.  We must also introduce at least one new scalar field $\mathcal{H}$ to break the new gauge symmetry.  The details of the breaking are only relevant to the DM phenomenology in two respects.  First, we need $\mathcal{H}$ to break $U(1)_{L_\mu - L_\tau}\to\set{Z}_3$, which requires that $\mathcal{H}$ have integer charge.  Second, we also require $\expect{\mathcal{H}} \sim m_\chi$, so that the annihilation $\chi\bar{\chi} \to Z'Z'$, where $Z'$ is the new gauge boson, is kinematically suppressed.  This in turn points to breaking by SM singlets.  We note that while \tabref{tab:LmLt} lists only a single new Higgs, it is possible that there could be additional scalars that contribute to the breaking, as in \refcite{1607.04046}, provided that they all have integer charge.

The dark sector charge assignment given in \tabref{tab:LmLt} is the unique one which allows the Yukawa couplings of \modeqref{eq:lagyuk}, as well as the scalar cubic $\Phi \, \Sigma^\dagger H$ necessary for dark partner decay.  It is thus the only possibility without adding further new particles.  However, it has the convenient feature that no further couplings are generated; in particular, the fact the tau multiplets have the same hypercharge but opposite $U(1)_{L_\mu - L_\tau}$ quantum numbers means the dark sector fields couple at tree-level \emph{only} to the second generation.  The scalar potential couplings $\mu_\Phi$ and $\lambda_3$ are likewise generated radiatively\footnote{$\rho_\Phi$ can be generated from a dimension-4 operator if there is a Higgs with $U(1)_{L_\mu - L_\tau}$ charge 2.}.  The accidental $\set{Z}_4$ symmetry discussed in \secref{sec:model} still exists if $\mu_\Sigma \to 0$, ensuring that radiative corrections to that coupling are proportional to it, \emph{i.e.} it is technically natural.